\documentclass[a4paper,fleqn,usenatbib, twocolumn]{aastex61}

\usepackage[T1]{fontenc}
\usepackage{ae,aecompl}

\usepackage{graphicx}	
\usepackage{url}        
\usepackage{multirow}   
\usepackage{amsmath}	
\usepackage{amssymb}	
\usepackage{textcomp}   
\usepackage{hyperref}   
\usepackage{booktabs}
\usepackage{dashrule}

\newcommand{\Hersc}{{\it Herschel}}
\newcommand{\hevics}{HeViCS}
\newcommand{\hatlas}{\textit{H}-ATLAS\/}

\newcommand\dg{$^\circ$}

\defcitealias{Valiante2016}{V16}

\received{July 3, 2017}
\accepted{November 14, 2017}
\submitjournal{ApJS}

\shorttitle{The \textit{Herschel}-ATLAS Data Release 2}
\shortauthors{Smith et al.}

\begin{document}

\title{The \textit{Herschel}-ATLAS Data Release 2, Paper I. Submillimetre and far-infrared images of the south and north Galactic Poles: The Largest \textit{Herschel} Survey of the Extragalactic Sky}

\correspondingauthor{Matthew W. L. Smith}
\email{Matthew.Smith@astro.cf.ac.uk}

\author[0000-0002-3532-6970]{Matthew W. L. Smith}
\affiliation{School of Physics \& Astronomy, Cardiff University, The Parade, Cardiff, CF24 3AA, UK.} 

\author{Edo Ibar}
\affiliation{Instituto de F\'isica y Astronom\'ia, Universidad de Valpara\'iso, Avda. Gran Breta\~na 1111, Valpara\'iso, Chile.}

\author{Steve J. Maddox}
\affiliation{School of Physics \& Astronomy, Cardiff University, The Parade, Cardiff, CF24 3AA, UK.}
\affiliation{Institute for Astronomy, The University of Edinburgh, Royal Observatory, Blackford Hill, Edinburgh, EH9 3HJ, UK.}

\author{Elisabetta Valiante}
\affiliation{School of Physics \& Astronomy, Cardiff University, The Parade, Cardiff, CF24 3AA, UK.}

\author{Loretta Dunne}
\affiliation{School of Physics \& Astronomy, Cardiff University, The Parade, Cardiff, CF24 3AA, UK.}
\affiliation{Institute for Astronomy, The University of Edinburgh, Royal Observatory, Blackford Hill, Edinburgh, EH9 3HJ, UK.}

\author{Stephen Eales}
\affiliation{School of Physics \& Astronomy, Cardiff University, The Parade, Cardiff, CF24 3AA, UK.}

\author{Simon Dye}
\affiliation{School of Physics and Astronomy, University of Nottingham, University Park, Nottingham, NG7 2RD, UK.}

\author{Christina Furlanetto}
\affiliation{School of Physics and Astronomy, University of Nottingham, University Park, Nottingham, NG7 2RD, UK.}

\author{Nathan Bourne}
\affiliation{Institute for Astronomy, The University of Edinburgh, Royal Observatory, Blackford Hill, Edinburgh, EH9 3HJ, UK.}

\author{Phil Cigan}
\affiliation{School of Physics \& Astronomy, Cardiff University, The Parade, Cardiff, CF24 3AA, UK.} 

\author{Rob J. Ivison}
\affiliation{Institute for Astronomy, The University of Edinburgh, Royal Observatory, Blackford Hill, Edinburgh, EH9 3HJ, UK.}
\affiliation{European Southern Observatory, Karl-Schwarzschild-Strasse 2, D-85748, Garching, Germany.}

\author{Haley Gomez}
\affiliation{School of Physics \& Astronomy, Cardiff University, The Parade, Cardiff, CF24 3AA, UK.}

\author{Daniel J. B. Smith}
\affiliation{Centre for Astrophysics Research, School of Physics Astronomy and Mathematics, University of Hertfordshire, College Lane, Hatfield, Hertfordshire, AL10 9AB, UK.}

\author{S\'{e}bastien Viaene}
\affiliation{Sterrenkundig Observatorium, Universiteit Gent, Krijgslaan 281, B-9000 Gent, Belgium.}
\affiliation{Centre for Astrophysics Research, School of Physics Astronomy and Mathematics, University of Hertfordshire, College Lane, Hatfield, Hertfordshire, AL10 9AB, UK.}

\begin{abstract}
We present the largest submillimeter images that have been made of the
extragalactic sky. The {\it Herschel} Astrophysical Terahertz Large Area Survey
(\hatlas) is a survey of 660 deg$^2$ with the PACS and SPIRE cameras 
in five photometric bands: 100, 160, 250, 350, and 500\micron. In this paper we present the images
from our two largest fields which account for $\sim$75\% of the survey.
The first field is 180.1 deg$^2$ in size centered on the North Galactic Pole (NGP)
and the second field is 317.6 deg$^2$ in size centered on the South Galactic Pole. 
The NGP field serendipitously contains the Coma cluster. 
Over most ($\sim$80\%) of the images, the pixel noise, including both instrumental
noise and confusion noise, is approximately 3.6, and 3.5\,mJy\,pix$^{-1}$ at 100 and 160\micron, and
11.0, 11.1 and 12.3\,mJy\,beam$^{-1}$ at 250, 350 and 500\micron, respectively, 
but reaches lower values in some parts of the images. If a matched filter is applied
to optimize point-source detection, our total 1$\sigma$ map sensitivity is
5.7, 6.0, and 7.3\,mJy at 250, 350, and 500\micron, respectively. 
We describe the results of an investigation of the noise properties of the images.
We make the most precise estimate of confusion in SPIRE maps to date finding values 
of $3.12 \pm 0.07$, $4.13 \pm 0.02$ and $4.45 \pm 0.04$\,mJy\,beam$^{-1}$ at 250, 350, and 500\micron\ in our un-convolved maps.
For PACS we find an estimate of the confusion noise in our fast-parallel observations of 4.23 and 4.62\,mJy\,beam$^{-1}$
at 100 and 160\micron. Finally, we give recipes for using these images to carry out
photometry, both for unresolved and extended sources.
\end{abstract}

\keywords{surveys - cosmology: observations - submillimetre: galaxies - galaxies: statistics - methods: data analysis}



\section{Introduction}
\label{sec:intro}

\begin{figure*}
  \centering
  \includegraphics[trim=0mm 141mm 0mm 40mm,clip=True, width=0.99\textwidth]{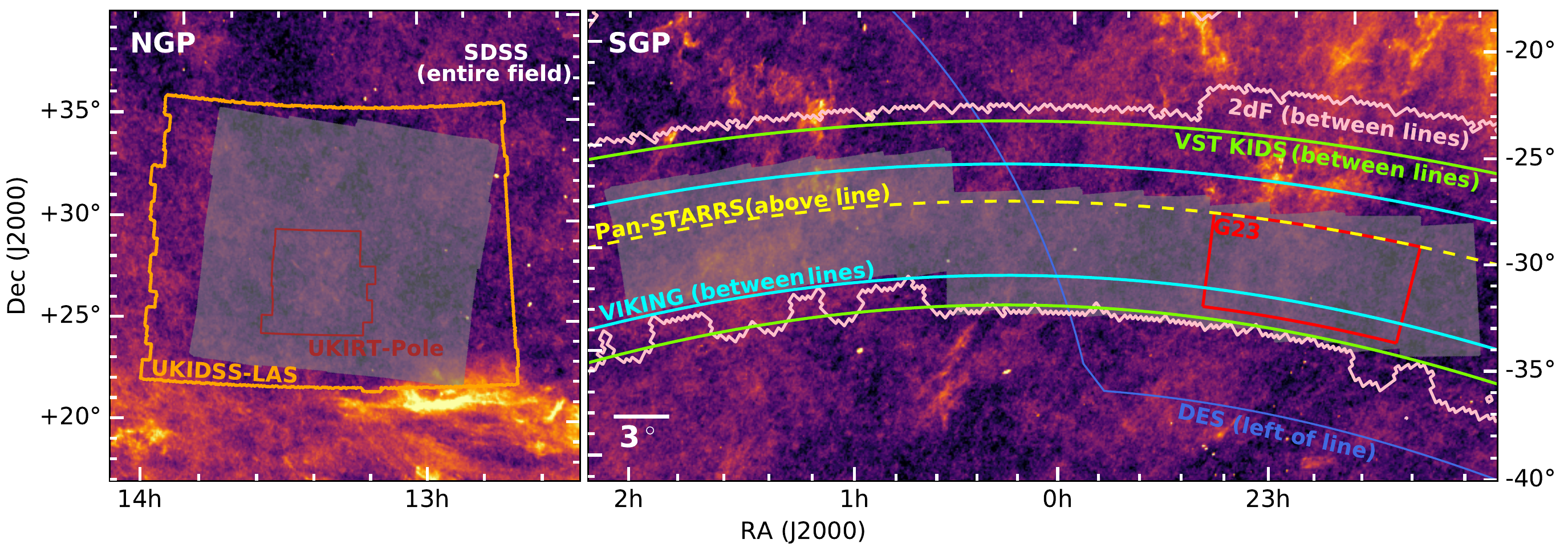}
  \figcaption{Overlapping, coverage of surveys in the NGP and SGP fields. The figure shows a 
              350\micron\ map from \textit{Planck} (color-image) for both fields (with the same angular scale), 
              with the gray regions illustrating the coverage of the \hatlas\ observations. The regions observed
              in complementary surveys are shown by the colored lines. In the NGP field, the entire \hatlas\ region 
              is observed by the UKIDSS-Large Area Survey and the SDSS, as well as the UKIRT-Pole survey which 
              covers 14\% of the \hatlas\ NGP field. In the 
              SGP there is overlapping coverage with the GAMA-23 hr field, VST-KIDS, VIKING, DES, Pan-STARRS and 2dF
              each with 18\%, 98\%, 79\%, 24\%, 50\% and 99\% overlap with \hatlas, respectively. For details of these surveys see
              Section~\ref{sec:intro}}
  \label{fig:surveyCoverage}
\end{figure*}

This is the first of three papers describing the second major data release of the 
{\it Herschel} Astrophysical Terahertz Large Area Survey ({\it Herschel}-ATLAS
or \hatlas), the largest single key project carried out
in open time with the {\it Herschel Space Observatory} \citep{Pilbratt2010}.
The \hatlas\ is a survey of approximately 660 deg$^2$ of sky
in five photometric bands: 100, 160, 250, 350, and 500\micron\ \citep{Eales2010}. 
Although the original goal of the survey was to study dust, and the newly formed
stars hidden by dust,
in galaxies in the nearby ($z<0.4$) universe \citep{Dunne2011,Eales2017}, in practice the 
exceptional sensitivity of {\it Herschel}, aided by
the large negative {\it k}-correction at submillimeter wavelengths
\citep{Blain1993}, has meant that the median redshift of the sources detected in the survey is $\simeq$1 \citep{Pearson2013}. 
The survey has therefore also already proved useful for astronomers
interested in studying galaxies in the early universe \citep[e.g.][]{Lapi2011} and also as a 
rich source of high-redshift galaxies, both objects that are lensed \citep{Negrello2010, Negrello2017, 
Gonzalez2012} and those that which are unlensed \citep{Ivison2016, Oteo2016}. The large area of the survey 
and the high Galactic latitude of the fields also mean that it can potentially be used to
look for Galactic objects with very low dust masses \citep{Eales2010,Thompson2010}.

The five \hatlas\ fields were selected to be areas with relatively little emission
from dust in the Milky Way, as judged from the IRAS 100\micron\ images \citep{Neugebauer1984}, and with a large
amount of data in other wavebands. In 2010 for the Science Demonstration Phase (SDP) of \Hersc, we released one 
16 deg$^2$ field in the GAMA 9 hr field \citep{Ibar2010,Pascale2011, Rigby2011, Smith2011}. 
Our first data release included three fields
on the celestial equator centred at approximately 9, 12, and 15 hr \citep{Bourne2016,Valiante2016}. 
These three fields, which cover 161 deg$^2$ constitute $\sim25$\% of the \hatlas\ survey, are 
rich in multi-wavelength data and in particular are covered by the Sloan Digital Sky Survey
\citep[SDSS;][]{Abazajian2009}, the VST Kilo-Degree Survey \citep[KIDS;][]{deJong2013},
the VISTA Kilo-Degree Infrared Galaxy Survey \citep[VIKING;][]{Edge2013},
the 2-Degree-Field Galaxy Redshift Survey \citep[2dF;][]{Colless2001},
and the Galaxy and Mass Assembly project \citep[GAMA;][]{Driver2009,Liske2015}. 
The data we released for these fields consisted of the {\it Herschel} images and
catalogs of the 120,230 {\it Herschel} sources and of 44,835 optical counterparts
to these sources.

Our second data release is for the two larger fields at the north and south Galactic
poles (NGP and SGP). The NGP field is centered approximately at R.A. of 13$^{h}$ 18$^{m}$ and
a decl. of +29$^{\circ}$\ 13$^{\prime}$ (J2000) and has
an area of 180.1 deg$^2$. The field is covered by the SDSS and has near-infrared
coverage from the UKIRT Infrared Deep Sky Survey Large Area 
Survey \citep[][]{Lawrence2007}. The \hatlas\ team itself also used UKIRT to carry out
a deep \textit{K}-band survey of part of the field (UKIRT Pole Survey), covering 25.93 deg$^2$ 
\citep[Paper III; ][]{Furlanetto2017}.
The NGP field contains the Coma cluster, and the {\it Herschel} images have been used to 
study the dust in the cluster galaxies \citep{Fuller2016}.

The SGP field is centered approximately at a R.A. of 0$^{h}$ 6$^{m}$ and a decl. of -32$^{\circ}$ 44$^{\prime}$ (J2000) 
and has an area of 317.6 deg$^2$. The field was covered by the 2dF spectroscopic survey and has been imaged
in four optical bands ($u$, $g$, $r$ and $i$) as part of KIDS, 
and in five near-infrared bands ($Z$, $Y$, $J$, $H$ and $K_s$) as part of the VIKING. 
The \hatlas\ data also cover the GAMA G23 field and has some overlap with
the Dark Energy Survey \citep{DES2016}, and Pan-STARRS \citep{Chambers2016}. Figure~\ref{fig:surveyCoverage} shows the regions
where complementary surveys overlap with the NGP and SGP fields.

Our data release for the \hatlas\ survey of the NGP and SGP is described in three papers. In
this paper, we describe the {\it Herschel} images and an investigation of
their statistical properties. We also give enough information for 
the astronomical community to be able to use these images to
carry out reliable photometry of individual objects and statistical `stacking' analyses
of classes of object. The second paper \citep[][]{Maddox2017} 
describes the catalogs of submillimeter sources found on the images. 
The third paper \citep{Furlanetto2017} describes a search for the optical/near-infrared counterparts
to the {\it Herschel} sources in the NGP field and the resulting multi-wavelength catalogue.
All the images described in this paper are available from
\url{www.h-atlas.org}, and Appendix~\ref{app:data} provides a guide to the products available, with a short description.

\section{Observing Strategy}
\label{sec:obsStrategy}

We observed the NGP and SGP using the same \Hersc\ observing mode as 
we used for the smaller fields on the celestial equator:
the SPIRE-PACS parallel mode in which both the SPIRE \citep{Griffin2010} and PACS \citep{Poglitsch2010} instruments are
used simultaneously. To maximise the area covered, and reduce potential 1/$f$ noise, 
we used the fastest scan speed of 60\,arcsec\,s$^{-1}$ (1/$f$ noise or ``low frequency noise" in bolometer timelines would lead to 
stripe artefacts in the map). 
Due to the offset between the cameras in the \Hersc\ focal plane, the PACS and SPIRE images are offset by
$\sim$22\arcmin, which means a tiny fraction (4\%) of both fields has data taken with only one camera.
We observed both fields at 100 and 160\micron\ with PACS and 250, 350, and 500\micron\ with SPIRE.

An observation consists of ``scan legs" where the telescope is moving at a constant velocity along a great circle across the field.
At the end of each scan leg, the telescope decelerates and then moves a constant distance
in an orthogonal direction to the beginning of the next scan leg, and then scans backwards across the field.
The total area covered by an observation is therefore built up by combining a large
number of scan legs during which the telescope is moving at a constant speed. 
Useful instrumental data is still being taken during the sections between scan legs and when the telescope
is accelerating, but in the \hatlas\ SPIRE maps these ``turn around'' data are not included in the final maps.
In parallel mode, the scan legs were separated by $155\,{\rm arcsec}$ in order to achieve a good coverage
with both PACS and SPIRE. More details can be found in the SPIRE and PACS Observers' Manuals, 
which are available at \url{http://herschel.esac.esa.int}.

For all \hatlas\ fields our observing strategy was to ensure all locations were
covered by two observations of each field
with roughly orthogonal scan directions. The scans needed to be roughly orthogonal 
because a major concern before launch was that
drifts in the bolometer signals of the instruments would
lead to artefacts on the images
with large angular scale. Obtaining observations with orthogonal scan directions makes it
possible, 
with the correct map-making algorithm,
to ensure that the final map does not contain any of these
artefacts
\citep{Waskett2007}.
In practice, SPIRE, although not PACS, proved sufficiently stable that 
it was possible to remove any drifts that did occur using information from
the thermistors attached to the bolometer arrays (Section~\ref{sec:SPIREtimelines}), and even maps made
from single observations were generally free of these artefacts. For PACS one of these advanced 
map-making algorithms is required; for details of the procedure we use see Section~\ref{sec:PACSdr}.

For the fields on the celestial equator, we followed this strategy by 
carrying out two observations with roughly orthogonal scan directions, each with an
exposure time of roughly nine hours and generally one after the other. A pair of observations
would cover a square area, or ``tile'', of side 4 degrees. Each of the equatorial fields was covered by four of
these tiles (\citealt{Valiante2016}, hereafter \citetalias{Valiante2016}). It was not possible to follow this
simple procedure for the NGP and the SGP because of the need to obtain uniform sensitivity
over such large fields and the sheer difficulty of scheduling such a large programme
during a three-year mission with all the geometric constraints on the pointing and scanning directions set by the
positions of the Sun and the Earth \citep{Waskett2007}.
Instead, for the NGP and SGP, we constructed the survey out of much bigger tiles, with
each tile being constructed out of two pairs of observations of rectangular regions of sky, with
the long axes (and scan directions) of the observations in each pair
being roughly parallel to each other and
roughly orthogonal to the long axes of the observations in the other pair.
The individual observations in the NGP typically had an observing time of $\sim$9.3--10.0\,hr.

\begin{figure*}
  \centering
  \includegraphics[trim=0mm 78mm 10mm 0mm,clip=True, width=0.99\textwidth]{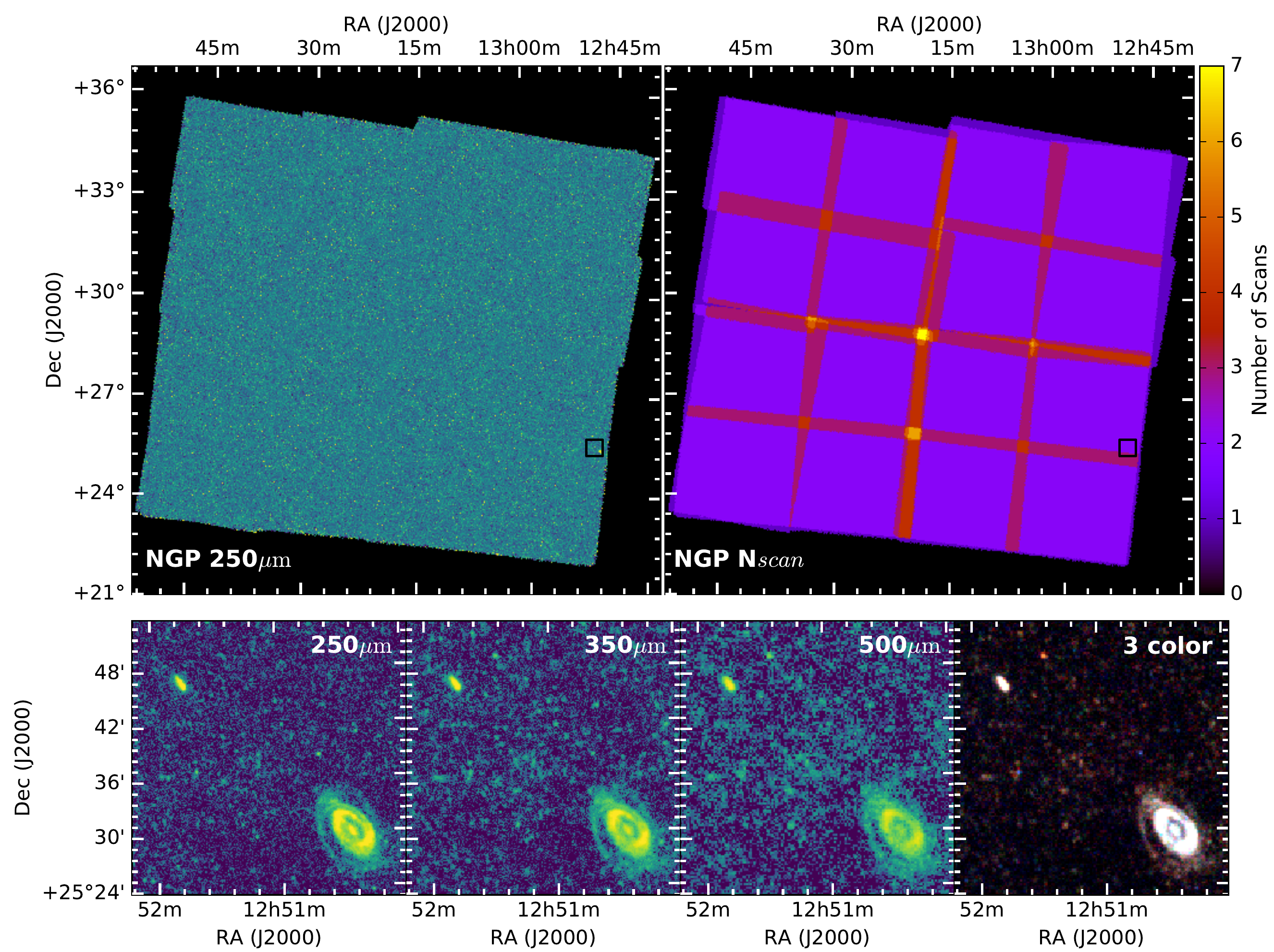}
  \figcaption{SPIRE maps of the NGP field. The larger top panels show the 250\micron\ map \textit{(left)}
           and the number of individual observations
           ($\rm N_{scan}$) from which the map was constructed \textit{(right)}. The row of
           lower panels show a close-up of a region centred on $\sim$13$^h$18$^m$,29$^\circ$18$^\prime$, which is 
           shown by the black rectangle in the top panels. The first three of the lower panels show the images at
           the three SPIRE wavelengths. The final panel shows an image formed by combining the three SPIRE images.
           On this three-color image, red shows sources that are brighter at 500\micron\ and
           blue shows sources that are brighter at 250\micron. Red sources will either be low-redshift
           galaxies with very cold dust or high-redshift galaxies. This region was not chosen at random and contains
           two local galaxies and a bright high-redshift lensed source (the red source at the top of the
           three-color image).}
  \label{fig:spireNGP}
\end{figure*}

\begin{figure*}
  \centering
  \includegraphics[trim=0mm 78mm 10mm 0mm,clip=True, width=0.99\textwidth]{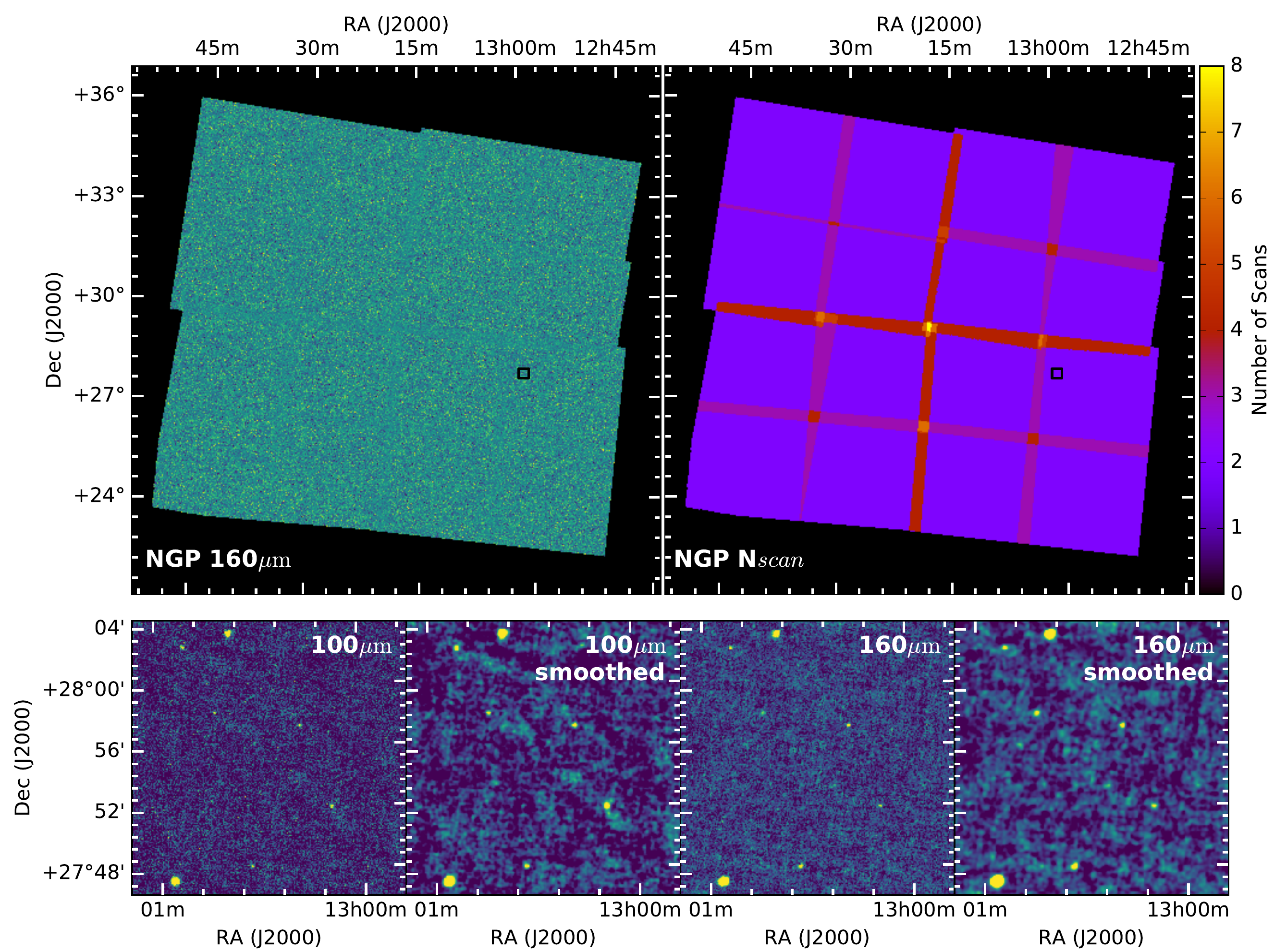}
  \figcaption{PACS maps of the NGP field. The larger top panels show the 160\micron\ map \textit{(left)}
           and the number of individual observations ($\rm N_{scan}$) from which the map was constructed \textit{(right)}.
           The row of lower panels show a small region of the Coma cluster at $\sim$13$^h$00$^m$, 27$^\circ$55$^\prime$ 
           (shown by the black rectangle on the top panels) at both
           PACS wavelengths, both the raw images and images smoothed by a Gaussian with same full-width at half maximum as
           the point-spread function.}
  \label{fig:pacsNGP}
\end{figure*}

The top right-hand panels of Figures~\ref{fig:spireNGP} and \ref{fig:pacsNGP} show our scanning 
strategy for the NGP, which was covered by four of these large tiles. Each tile is almost a square with
sides of 
$\simeq$7.2\dg and $\simeq$6.5\dg.
Given the
scheduling constraints, it was not possible to make all the tiles line up precisely, and
to ensure complete coverage of each field, we made the tiles overlap slightly. The 
entire area covered by our observations is roughly a rectangle
with dimensions of $\simeq$14.0\dg
by $\simeq$12.8\dg. 
The area of the field with useful data is 180.1 deg$^2$.
The figures
show the number of observations at each point for SPIRE (Figure~\ref{fig:spireNGP}) and PACS (Figure~\ref{fig:pacsNGP}). 
They show that the design of our observing programme was
quite successful,
since
most of the NGP has data from two roughly orthogonal
observations, with narrow strips having data from four observations (and thus an increase 
in sensitivity), and with a few very small areas having data from even more observations
and thus even better sensitivity.

\begin{figure*}
  \centering
  \includegraphics[trim=0mm 78mm 10mm 0mm,clip=True, width=0.99\textwidth]{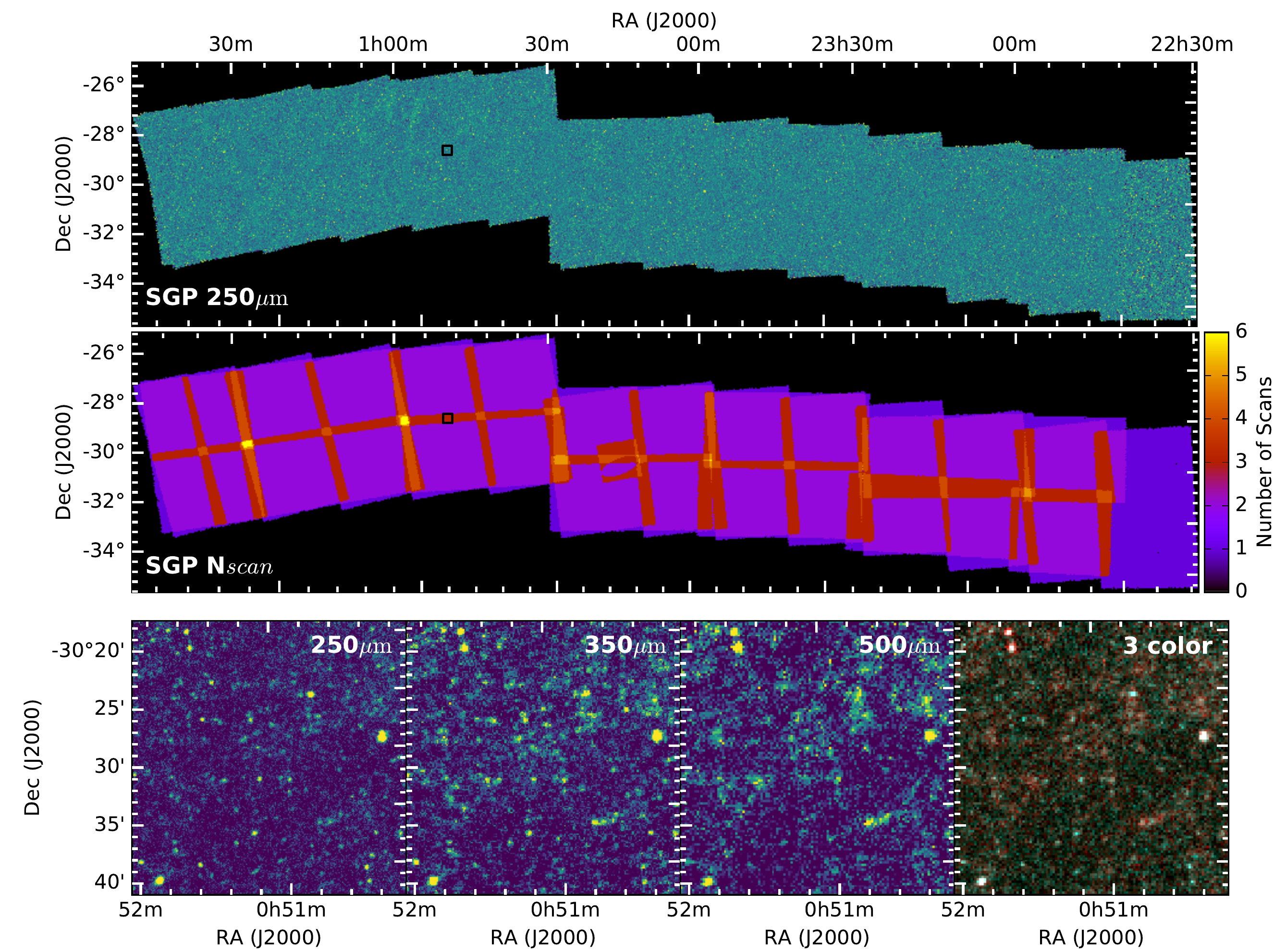}
  \figcaption{SPIRE maps of the SGP field. The larger top panels show the 250\micron\ map \textit{(top)}
           and the number of observations ($\rm N_{scan}$) from which the map
           was made \textit{(middle)}. Note the region at the western edge of the field where the map was made
           from only a single observation (see the text). The row of
           lower panels show a close-up of a region centered on $\sim$ 0$^h$51$^m$, -30$^\circ$30$^\prime$,
           which is shown by the black rectangle on the top panels.
           The first three of the lower panels show the images at the
           at the three SPIRE wavelengths. The final panel shows an image formed by combining the three SPIRE images.
           On this three-color image, red shows sources that are brighter at 500\micron\ and
           blue shows sources that are brighter at 250\micron. Red sources will either be low-redshift
           galaxies with very cold dust or high-redshift galaxies.}
  \label{fig:spireSGP}
\end{figure*}

\begin{figure*}
  \centering
  \includegraphics[trim=0mm 78mm 10mm 0mm,clip=True, width=0.99\textwidth]{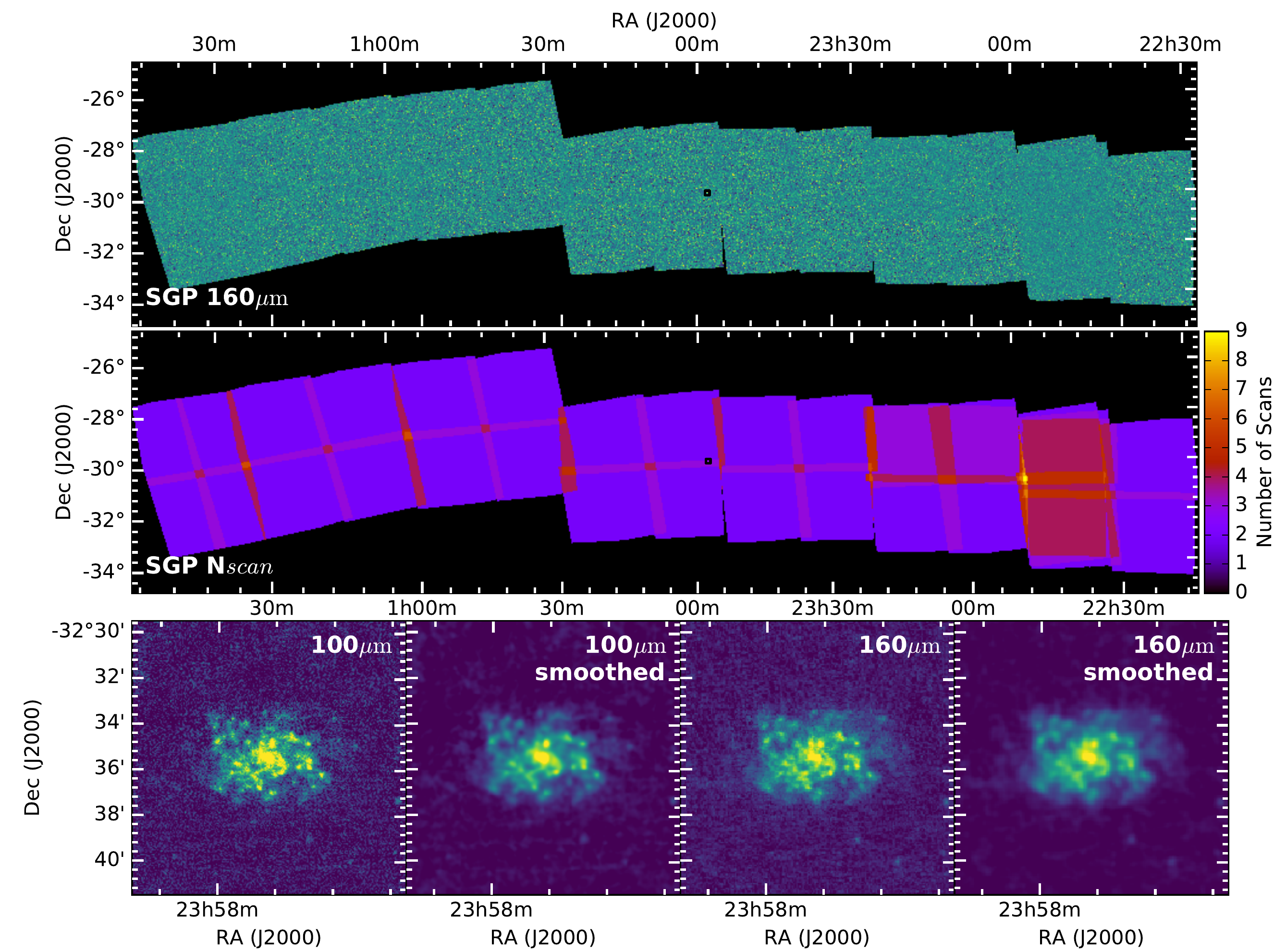}
  \figcaption{PACS maps of the SGP field. The larger top panels show the 160\micron\ map \textit{(top)}
           and the number of observations ($\rm N_{scan}$) from which the map was made \textit{(middle)}.
           Note the difference from the coverage maps shown in Figures~\ref{fig:spireSGP}; there are several
           regions where there are extra data for PACS because of parallel-mode observations
           performed to replace failed SPIRE observations, and there is also the area at the western
           edge of the field where we were unable to get a replacement for a failed SPIRE observation
           (Figure~\ref{fig:spireSGP}) but the PACS observation was fine. The row of lower panels show close-up images
           of the area around the nearby galaxy NGC\,7793, which is shown by 
           the black rectangle on the top panels, at both PACS wavelengths, both the
           raw images and images smoothed by a Gaussian with the same full-width at half maximum as the point-spread function.}
  \label{fig:pacsSGP}
\end{figure*}

For the SGP, we adopted the same procedure of creating roughly square tiles out of 
two pairs of parallel rectangles. The design of the survey is shown in
the center panels of Figures~\ref{fig:spireSGP} and \ref{fig:pacsSGP}, which show the coverage of SPIRE
and PACS, respectively. The tiles form two rough rectangles which are touching
but offset from each other. The shape of
the SGP field is different from the one we envisioned before launch \citep{Eales2010};
the new design maximizes the overlap with the 2dF spectroscopic survey and
the new spectroscopic survey carried out by the GAMA team at an R.A. of $\simeq$23$^h$. 
The area of the field with useful data is 317.6 deg$^2$.
The individual observations in the SGP had a typical exposure time of $\sim$9.3--10.1\,hr.

The shape and size of the SGP field means that the tiles
do not line up so well as for the NGP, therefore the coverage is slightly less uniform
(Figures~\ref{fig:spireSGP} and \ref{fig:pacsSGP}).
The coverage was also less uniform due to two complications. The first was that
during {\it Herschel} observation 1342196626 a planet (either Jupiter or Uranus) 
was at a position where light from it was reflected by the support structure of the secondary mirror
into the SPIRE instrument, leading to a `stray-light' feature on the image. After we discovered this feature, the
\textit{Herschel Science Centre} scheduled a replacement observation (obsid: 1342245911) covering
an area $\simeq$1.8\dg$\times$1.7\dg in size to patch the image. The patch can be seen in Figure~\ref{fig:spireSGP} at 
R.A. $\sim 0^h 16^m$, decl. $\simeq-32^\circ 43^\prime$.

The second complication is that occasionally during our observations the SPIRE instrument
went into safe mode, probably because it was hit by a cosmic ray, while PACS kept on
observing. As a result, there is a region ($\sim$6.0\dg$\times$3.5\dg) 
at the western end of the SGP for which we have only one observation at each point for SPIRE (Figure~\ref{fig:spireSGP}) but the
normal coverage with PACS (Figure~\ref{fig:pacsSGP}); we were not able to obtain a replacement
SPIRE observation because we ran out of allocated observing time.
There are also two regions where we did succeed in getting replacement observations with SPIRE,
and as a result we have better than the usual coverage with PACS.
These regions are both toward the western end of the image (Figure~\ref{fig:pacsSGP}).
One is $\sim$6.0\dg$\times$3.5\dg in area, for which at most points
we have four observations rather than the usual two, and the other
is a region $\sim$6.2\dg$\times$3.5\dg in size, for which
at most points we have three observations rather than the usual two.

\section{The SPIRE Observations}

\hatlas\ imaged the sky with the SPIRE camera simultaneously through
three submillimeter filters centered at 250, 350 and 500\micron. Each
filter was approximately 30\% wide in $\Delta \lambda / \lambda$.
A full description of the instrument is in \citet{Griffin2010}.
We have given a very detailed description of the 
data reduction below in order to make clear the differences
in the procedure for the NGP and SGP fields to those used for the fields on the celestial
equator \citepalias{Valiante2016} and from the procedures used
for other {\it Herschel} surveys.

In Section~\ref{sec:SPIREtimelines} we describe our treatment
of the SPIRE timelines from the raw data to reduced timelines. In 
Section~\ref{sec:SPIREastrom} we discuss how we correct the astrometry in
each tile and our iterative technique to remove glitches. Section~\ref{sec:SPIREmaps}
describes the final map products, our method to remove contaminating 
emission from Galactic cirrus and filtering applied to optimize detection
of point sources. Finally, in Section~\ref{sec:SPIREcal} we describe the
calibration and the differences from the calibration used for the GAMA fields.

\subsection{The SPIRE Bolometer Data} 
\label{sec:SPIREtimelines}

The SPIRE instrument consists of three imaging arrays for
observations at 250, 350, and 500\micron\ with 139, 88, and 43 bolometers,
respectively. Each array has two associated thermistors to monitor the
temperature of the array, although after launch only one of the 350\micron\ thermistors
worked, and two dark bolometers, bolometers that receive no light.
In {\it Herschel} parlance, the `level-0' data are the raw instrumental/telescope data
and the `level-1' data consist of the calibrated flux-density versus time
measurements for the individual bolometers (calibrated timelines), which can then
be used to create an image of the sky. In this section we explain the way we produced the calibrated
level-1 data.

We converted the level-0 data to the level-1 data, the calibrated timelines,
using the \Hersc\ Interactive Processing Environment \citep[HIPE,][]{Ott2010}, 
version 11.0 1200 (development build).  
Unless described otherwise, we used the standard components
of the data-reduction pipeline. Forty two of 51 observations
for the NGP and SGP fell in observing days 320--761, during which
there were positional offsets caused by a change in the operating temperature of the
star-tracker camera for which the camera was not re-calibrated. We used an
updated pointing model released by the \textit{Herschel Science Centre} in 2012
to correct the pointing for these observations.

We corrected glitches in the bolometer and thermistor data
using a different technique from the standard module in the pipeline.
Instead of the default wavelet deglitcher, we used the sigma-kappa deglitcher,
since tests on both our \hatlas\ data and on data from
the {\it Herschel} Virgo Infrared Cluster Survey \citep[\hevics,][]{Davies2010} showed it performed better
for parallel-mode observations with high scan-speeds and a reduced (10\,Hz) sampling rate (the
non-default settings for the \textsc{sigmaKappaDeglitcher} task were: kappa=3.5, gamma=0.1, 
boxFilterCascade=5, largeGlitchRemovalTimeConstant=7, iterationNumber=3).

At this point, we had calibrated timeline data (level-1) but the data were still affected
by artefacts, including `jumps' and gradual changes in the signal caused by changes
in the temperature of the bolometers. The thermistor timeline data contained the necessary
information to correct for the effect of temperature. Since the 350\micron\ array only had one
working thermistor, we used dark-bolometer 1 as a replacement. However, before we corrected
the bolometer timelines, it was necessary to correct both the bolometer and thermistor
timelines for the jumps.

\begin{figure}
  \centering
  \includegraphics[trim=0mm 3mm 0mm 0mm,clip=True, width=0.49\textwidth]{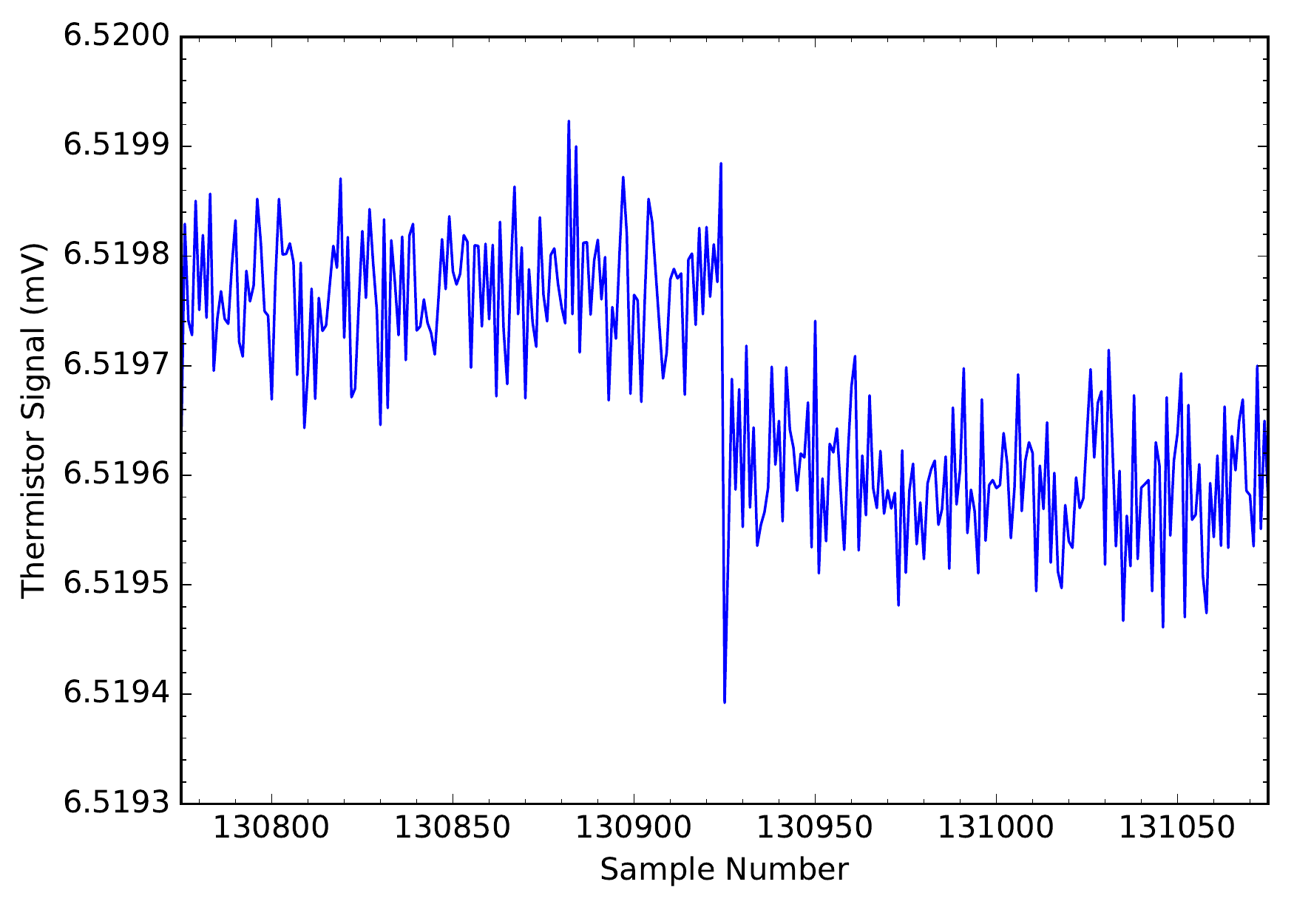}
  \figcaption{Example of a thermistor `jump' from one of the thermistors
for the 500\micron\ array, PLWT2, during ATLAS observation 1342196626.}
  \label{fig:jump}
\end{figure}

A jump is an instantaneous
change in the voltage of an individual bolometer or thermistor (a typical example is shown in Figure~\ref{fig:jump}). In rare
cases, rather than a step change in voltage, there is a sudden large change
followed by a gradual decay back toward the original value. 
Jumps appear to be more common in thermistors than bolometers.
Our jump correction method (see below) does not work well in correcting these rare jumps, and
if one of these occurred in a thermistor, we used one of the dark bolometers as a replacement when
correcting the bolometer timelines for the effect of temperature.
We looked for jumps in bolometers and thermistors in different ways.

For the thermistors, rather than using the automatic jump detector, we 
inspected both thermistor timelines for each array by eye
to spot jumps in the timelines, using the Kst visualization tool\footnote{Kst is a data visualization tool. For 
more information see \url{http://kst-plot.kde.org/}.} In the case of the 350\micron\ array, we
carried out a similar inspection of the timelines of dark-bolometer 1.

It was not practical to search for jumps in the bolometer timelines in the same way because
there were too many of them. Instead we made initial maps of each individual observation
from the timelines and visually searched for the light and dark thin streaks caused by
jumps; since a single \hatlas\ observation consists of scans in a single direction and one map pixel usually
only includes data from a few bolometers, the effects of a jump are easy to see.

Before correcting the jumps, we combined
all scan legs from an observation, including the ``turn-around'' data, into a single timeline.
We then corrected all the jumps in the timeline by fitting a linear relationship to portions of the timeline 
immediately before and after the jump, and then adding the difference in these relationships to the timeline after the jump 
\citep[see][for more details]{Smith2012}.
We replaced the samples immediately around the jump (Figure~\ref{fig:jump}) with random
noise, and these samples were then masked and not used to make the final maps.

The advantage of combining the data from all scan legs into
a single timeline is that makes it possible to remove more accurately the
drift in the bolometer signals caused by temperature changes. In the standard pipeline, this correction is done
separately for each scan leg and the information in the ``turn-around'' data is not used at all.
Before we made the correction for the effect
of temperature, we masked any samples in the timelines that had been flagged as bad 
(e.g., samples effected by glitches, samples in which the signal is saturated) and any places in the timelines
where there were obvious bright sources.

We fitted the following relationship between the signal measured by a single thermistor ($S_T$) and the signal measured by the
$i$th bolometer ($S_{{\rm bolom},i}$): $S_{{\rm bolom},i} = a \times S_T + c$. We then subtracted this relationship from the bolometer signal,
effectively removing the effect of the temperature change. A difference from our procedure in the GAMA fields
\citepalias{Valiante2016} is that we carried out this fit for both thermistors, and then used the thermistor that
produced the best fit to the data
to correct the bolometer timelines. For the 350\micron\ timelines, we did the fits
for the one working thermistor and for dark-bolometer 1. For any 
observation for which one of the thermistors was saturated or affected by a jump that could not be accurately
corrected (see above),
we used one of the dark bolometers rather than the thermistor. There were parts of some timelines where 
the linear relationship given above
did not provide a good fit to the
data. These almost always occurred six hours after a cycle of the SPIRE cooling system and became
known as ``cooler burps''. In cooler burp regions the timelines vary far more more rapidly than the average SPIRE timeline. 
For these timelines we fitted a fifth-order polynomial rather than a linear relationship.

Once the thermal drift correction had been applied, we applied a high-pass filter to remove any residual drifts. 
Before applying the filter, we removed the brightest sources from the timelines, 
and then restored these after the filtering. The high-pass filter corresponds to a scale on the
sky of 4.2\dg, which was chosen to minimize the 1/$f$ noise on
the images \citep{Pascale2011}. Our images will therefore not contain any structures on scales larger than this.
In practice, however, one of the effects of our scanning method is that any structure on scales $>$20\arcmin\ is
attenuated \citep[][see Section~\ref{sec:powerSpec} for more details]{Waskett2007}.

\subsection{Initial Maps and Astrometry}
\label{sec:SPIREastrom}

The next step in the data reduction was to make maps from each individual observation. The two purposes of these initial maps
were to check the astrometry of each observation and to remove more bad data, since 
some low-level artefacts and samples containing bad data are easier to find on the maps than in the timelines.

We carried out our astrometric calibration of the observations in the NGP using the technique that is
described more fully in \citet{Smith2011}. Briefly, we produced initial source catalogs for each
map using our source-detection method \citepalias{Valiante2016}. We then produced histograms of the differences in 
RA and Dec between the positions of the sources 
and the positions of all objects on the SDSS DR7 \textit{r}-band 
images \citep{Abazajian2009} within 50\arcsec\ of
each source. We then fitted these distributions using a Gaussian model for the SPIRE positional errors,
allowing for the effect of clustering in the SDSS data \citep[see][for the details]{Smith2011}.
This procedure allowed us to measure
the average difference in positions in both R.A. and decl. for each dataset between the {\it Herschel} positions
and the SDSS positions with a precision of $\sim$0.05 arcsec in each direction. The shifts
we found ranged from less than an arcsec to a few arcsec, in agreement with the 1$\sigma$ pointing
uncertainty of $\sim$2 arcsec given for {\it Herschel} \citep{Pilbratt2010}.
We used these shifts to correct the astromometry for each {\it Herschel} observation, so that the
effective calibration of the NGP maps and catalogs should be the same as the SDSS.

The SGP field is not covered by the SDSS. To calibrate the astrometry of the SGP observations, we used the same method
as above but we replaced the SDSS catalogue with the catalogue from the VLT Survey Telescope ATLAS \citep{Shanks2015}.
We found astrometric shifts of a similar size to those for the NGP except a few tiles with corrections of $\sim$4 arcsec.

We also used these initial single-observation maps to look for any residual artefacts. In particular,
the standard deglitcher modules (Section~\ref{sec:SPIREtimelines}) do not completely mask very large glitches,
due to a parameter which limits the maximum number of samples that can be masked. These unmasked samples 
produce `glitch tails' on the images, linear features in the scan direction.
We looked for these by eye on the map and then masked the appropriate parts of the timelines.

We also used a new iterative technique to look for glitches that were too faint to be detected by algorithms that work on the timelines.
This technique was not applied to the data for the GAMA fields, since we developed it after our first data release. 
In this technique, we look for bolometer samples that are discrepant by at least 5$\sigma$ from the value predicted
from the statistics of the map. The first step in the procedure is to use all the observations to make a low-resolution
map of each field with three times the default pixel size (Section~\ref{sec:SPIREmaps}).
The flux in each pixel, $F_{\rm MapPixel}$, is the mean of the fluxes of the $N_{\rm Pixel}$ timeline samples
that fall within that pixel. If $F_{\rm Sample}$ is the flux density of a single bolometer sample and $\sigma_{\rm MapPixel}$
is the error in $F_{\rm MapPixel}$, we treat a sample as bad, and thus mask it, if

\begin{equation}
\label{equ:2deglitch}
\frac{F_{\rm Sample} - F_{\rm Map Pixel}}{\sigma_{\rm Map Pixel} \sqrt{N_{\rm Pixel}}} > 5
\end{equation}

\noindent After masking all the discrepant samples in the
timelines, we remade the maps and looked for additional samples
that met the criterion in equation~\ref{equ:2deglitch}. We carried out four iterations
of this procedure, masking in total 295,496 and 539,126 bolometer samples  for the NGP and SGP, respectively, which
equates to $\sim$0.02\% of all bolometer samples.

As the final step in the processing of the timelines, we masked the ``turn around'' data (i.e., regions where the satellite
was not scanning at a constant speed) at the end of each scan leg.

\subsection{The Final SPIRE Maps}
\label{sec:SPIREmaps}

We created the final maps by combining all the corrected and masked level-1 data for each field. We used the simple
(`naive') map-maker in which the flux density in each pixel is taken to be the mean flux density of all the bolometer
samples that contribute to that pixel. Despite the concern before launch \citep[Section~\ref{sec:obsStrategy};][]{Waskett2007}
that sophisticated map-making algorithms would be necessary to remove large-scale artefacts, the use
of the thermistors to correct for the thermal drift of the arrays worked well enough that this simplest of
all map-making techniques was sufficient. As for the GAMA fields \citepalias{Valiante2016}, we used a pixel
size of 6\arcsec, 8\arcsec, 12\arcsec\ for the 250, 350, and 500\micron\ bands, respectively, which are different from the 
default pixel sizes of {\it Herschel} images; we chose them because they correspond to roughly one third of the 
size of the PSF (full-width half maximum; FWHM) in each band (see below) and they are big enough that the chance
of a map pixel containing no bolometer samples, thus producing a Not-a-Number (NaN) pixel, is low.
The 250\micron\ maps for both fields are shown in Figures~\ref{fig:spireNGP}~and~\ref{fig:spireSGP}.

The standard pipeline produces an estimate of the uncertainty in the flux density measured in each pixel by calculating the
variance of all the timeline samples that contribute to that pixel. However, this method does not produce an accurate
estimate of the flux uncertainty for two reasons: (a) the small number of samples
in each pixel means that the error on the uncertainty estimate is quite large; (b) the variance will be too high if the pixel
coincides with a bright object. Instead we have produced our own uncertainty maps, using $\sigma_{\rm inst}/\sqrt{N_{\rm sample}}$  
as our estimate
of the uncertainty in flux density, in which $\sigma_{\rm inst}$ is the instrumental noise for one timeline sample, calculated
using the method described in Section~\ref{sec:SPIREinst}, and $N_{\rm sample}$ 
is the number of timeline samples contributing to that pixel.
This, of course, is an estimate of the uncertainty in flux density arising from instrumental noise and does not
include the effect of source confusion (Section~\ref{sec:SPIREconf}).

For our PSF, we use the same PSF that was determined by \citetalias{Valiante2016} from images of Neptune
with the same pixel size as the \hatlas\ images (see that paper for more details). The FWHM
of the azimuthally averaged PSF is 17.8, 24.0, and 35.2 arcsec at 250, 350, and 500\micron, respectively.

As part of the data release, we have also produced images optimized for the detection of point sources.
The first step in producing these images was to remove any large-scale structure from the images,
which is mostly emission from Galactic dust (`cirrus' emission). We removed the cirrus emission using the 
\textsc{Nebuliser}\footnote{\url{http://casu.ast.cam.ac.uk/surveys-projects/software-release/background-filtering}}  algorithm
developed by the Cambridge Astronomical Survey Unit and we refer the interested reader to \citetalias{Valiante2016}
for the details of how we did this. The result of the application of {\sc Nebulizer} is that the images should not contain any
emission, whether from Galactic dust or from outside the Galaxy, with an angular scale $\succeq$3 arcmin.

For an image containing only one point source and instrumental noise, the maximum signal-to-noise for
the source is obtained by convolving the image with the PSF \citep{North1943}. However, the noise in the \hatlas\ images
is a combination of instrumental noise and ``confusion noise'', the result of the large number 
of submillimeter sources that are too faint to detect individually but which merge together
to form an undulating background to the images. \citet{Chapin2011}
have shown how to calculate a convolving function or `matched filter'
that will produce the maximum signal-to-noise for an unresolved source for any ratio of confusion to instrumental
noise. As part of this data release, we have produced images optimized for finding point sources by
convolving the raw images with the matched filters we used for the GAMA fields \citepalias{Valiante2016}. 
The instrumental noise and confusion noise for the SGP and NGP are actually slightly different
than for the GAMA fields (see Section~\ref{sec:SPIREinst} and \ref{sec:SPIREconf}), 
which means that the matched filters we have used are not precisely optimized, 
but this small disadvantage is outweighed by our being able to use the results of the extensive simulations we
carried out with the GAMA matched filters \citepalias{Valiante2016}. We tested the effect of using a matched filter optimized
with our updated noise values and found that the difference in the number and fluxes of sources is negligible.
For readers interested in measuring the flux density of a point source,
the matched-filtered images are the ones to use.


\subsection{The SPIRE Flux Calibration}
\label{sec:SPIREcal}

The flux calibration we applied to the images was publicly released as calTree v11. 
There is no difference between this and the most recent (at the time
of writing) flux calibration (calTree v14).
Because of the change in the SPIRE calibration, the flux densities
for the NGP and SGP are not quite on the same scale as
those in the GAMA fields \citepalias{Valiante2016}.
To create the maps described by \citetalias{Valiante2016}, we
used the SPIRE v5 calibration tree to create
the level-1 data, but applied a 1.0067 correction factor to the 350\micron\ data to make the
effective calibration the same as
calTree v8. Between v8 and v14 the average (multiplicative) change in flux density is
1.0253$\pm$0.0012, 1.0182$\pm$0.0045, and 
1.0125$\pm$0.0006 at 250, 350, and 500 $\mu$m, respectively.
Therefore, to put the \hatlas\ GAMA flux densities on the same scale as those
for the NGP and SGP, we need to multiply the flux densities from the DR1 release by these factors.
Note, however, that because each
bolometer is calibrated individually, the actual correction factor for an individual source 
depends on which individual bolometers crossed that position.

\section{The PACS Observations}
\label{sec:PACSdr}

We observed the sky simultaneously at 100 and 160\micron, using the PACS camera
\citep{Poglitsch2010}. While PACS also has a photometric band at 70\micron, it cannot
observe in both the 70 and 100\micron\ bands at the same time, and we chose to observe
in the 100\micron\ band.
The passband filters are relatively broad with $\bigtriangleup\lambda/\lambda \sim 1/3$ for both
wavelengths; the detailed filter response curves can be found in the HIPE calibration product 
and are shown in the PACS Observer's Manual\footnote{\url{http://herschel.esac.esa.int/Docs/PACS/pdf/pacs\_om.pdf}}.
Due to the offset between the SPIRE and PACS instruments in the focal plane 
of the telescope, there is a $\simeq$22 arcmin offset
in the final PACS and SPIRE images.

The PACS datasets were more challenging to reduce than the SPIRE dataset
because they were larger in volume and because the noise power
on the PACS images has a weak dependence on spatial
frequency ($\propto 1/f^{\alpha}$ with $\alpha\simeq0.5$), which makes
it impossible to reduce the noise by spatial filtering
without affecting the properties of extended sources.
The PACS datasets for the NGP and SGP were even larger than for the
GAMA fields, because each tile is constructed from at least four
observations (Figures~\ref{fig:pacsNGP} and \ref{fig:pacsSGP}) rather than the
two used to make the GAMA tiles.

We processed the PACS data up to the stage of the calibrated timelines (level-1 data)
in exactly the same way as described by \citetalias{Valiante2016} for the GAMA fields and we refer the reader to that
paper for the details.

We calibrated the astrometry of each observation using a different method from the
one we used for the GAMA fields. 
For the latter we measured the positions on the PACS images
of sources also detected on the SPIRE images, and thus tied the
PACS astrometry to the SPIRE astrometry and ultimately to the SDSS astrometry. For the SGP and NGP fields
we used a different approach. We first made a `naive' map from each
individual observation, in which the flux density
in each pixel is
estimated from the average of the timeline samples falling in that
pixel. We then found all the 3.4\micron\ sources from the \textit{WISE} survey \citep{Wright2010} 
that fell within the area covered by the map. Next we extracted small parts of the PACS image centered on each \textit{WISE} source and added these `cutouts' together
to produce an average PACS source. Finally, we measured the offset 
between the peak of the PACS emission and the expected
position 
from the \textit{WISE} astrometry. We found offsets between 0.2 and 2.0 arcsec.
Before making the final maps, we corrected the astrometry of each individual observation 
using these offsets.

The effect of `1/\textit{f}' noise (see above) is that naive maps made from the PACS data are
dominated by noise on large angular scales unless strong filtering is applied. For the GAMA fields, we tested a number of
more sophisticated map-making
techniques, eventually choosing the {\it Jscanamorphos} algorithm \citep{GraciaCarpio2015}, a version
installed as part of HIPE 
of the {\it Scanamorphos} algorithm \citep{Roussel2013}.
We decided to use this algorithm for the SGP and NGP, but then encountered the complication
that {\it Jscanamorphos} could only make a map from two orthogonal observations. If more than 
two observations are needed, a map is made for each pair and then all maps are averaged together. 
We adapted the standard HIPE script for
{\it Jscanamorphos} (from developers build 13.0.5130) so that it
would allow us to use all four observations simultaneously\footnote{The script is available on GitHub 
\url{https://github.com/mwls/Public-Scripts}.},
a necessary requirement to make one of the NGP and SGP tiles due to the scanning strategy.
We found no detrimental effects on the PACS images from combining our individual observations with slightly 
different scan angles or from combining data taken on different observing days. Despite modifying the script to use as
little memory as possible and running on a 158\,GB RAM machine, the 100\micron\ 
data of the westernmost field of the SGP with seven observations 
(instead of the usual four, see Figure \ref{fig:pacsSGP}) could not be processed in one {\it Jscanamorphos} process.
In this one field, we separated the observations into two (each had a coverage of at least two observations), 
and made tile maps out of each set of data.

We removed residual large-scale 1/$f$ noise from the {\it Jscanamorphos} map of each tile by applying \textsc{Nebuliser}.
This applies an iterative sliding median and linear filter to remove 
large-scale structure in an image. We set the filter to remove emission on scales
above 300\arcsec\ for both the 100 and 160\micron\ bands. We chose this value to
preserve the flux from galaxies smaller than $\simeq$100\arcsec\ in radius, which is true
of all but a few of the biggest galaxies in the fields (for these the flux densities
can be measured from the raw \textit{Jscanamorphos} maps). 
After the application of \textsc{Nebuliser}, we cropped each map to an area
covered by at least two orthogonal observations, which ensured
that the final images should have no large-scale artefacts caused by the 1/$f$ PACS noise.

We applied SWarp\footnote{\url{http://astromatic.iap.fr/software/swarp}} 
\citep[v2.19.1,][]{Bertin2002} to mosaic the individual
tiles and create the final maps for this data release.
These images have a pixel size of 3 and 4\arcsec\ at 100 and 160\micron\ band, respectively,
which is roughly one third of the size of the PACS PSF
(FWHM).
We have also provided, as part of the data release, images showing the number of
observations ($\rm N_{scan}$) contributing to the flux density in each pixel.
Figures~\ref{fig:pacsNGP} and \ref{fig:pacsSGP} show the PACS 160\micron\ images and the $\rm N_{scan}$
images for both fields. The lack of any regions with $\rm N_{scan} = 1$ is
because of the requirement that there be at least two roughly orthogonal scans contributing to each pixel.

The PACS PSF depends on the observing mode, the pixel size in the map, the spectral energy distribution (SED) of the
source, and the algorithm used to make the map \citep{Lutz2015}. A particular problem
is that in parallel mode with fast-scanning (60\arcsec/s) the PSF is elongated in
the scan direction, especially at 100\micron, because of the on-board averaging of the PACS data necessary to
transmit both the PACS and SPIRE data to Earth. For the GAMA fields we developed a method of constructing
an empirical PSF from the data themselves \citepalias{Valiante2016}. It was not possible to use this method on
the NGP and SGP fields because they were not covered by the GAMA survey, 
so we have simply assumed our analysis of the PSF in the GAMA fields can be used for the NGP and SGP, 
as observing mode, pixel sizes, and mapping algorithm are almost identical.

We fitted an azimuthally symmetric Gaussian to the empirical PSF, obtaining a value
for the FWHM of 11.4 and 13.7 arcsec at 100 and 160\micron, respectively. Exact PSFs could be calculated
\citep[e.g.,][]{Bocchio2016} based on the scanning angle of the observations, but due to the large number of
combinations of observations, we considered this impractical.
We recommend that anyone wishing to convolve the images should use these Gaussians rather than the
empirical PSF, which we have not released because our method for constructing the empirical PSF
leads to some systematic uncertainty in the values
of the central pixels.

For those interested in aperture photometry, we have provided as part of the data release a table
listing the encircled energy fraction (EEF) of the PSF against radius.
This is derived from our empirical PSF for radii less than 30 arcsec and from the EEF produced by
the PACS team for radii between 30 and 1000 arcsec; we refer the reader to \citetalias{Valiante2016} 
for how this was done.

\section{Photometry on the SPIRE Images}

In this section we describe an investigation of the characteristics of the SPIRE images and give the reader
the information necessary to carry out photometry on the images, both of point sources and extended sources.
We first describe an investigation of the instrumental noise and the confusion noise, which
both make a significant contribution to the total noise on the images.

\subsection{The SPIRE Instrumental Noise}
\label{sec:SPIREinst}

In determining the instrumental noise, the first step is to
remove any real astronomical signal (e.g. galaxies, cirrus, confusion noise) 
by creating a jackknife map from subtracting two images of the same part of the sky made
from individual SPIRE observations. The instrumental noise can then be measured from the jackknife map.
For all pixels in both the NGP and SGP that are covered by at least two individual observations, we 
calculated the instrumental noise per single bolometer sample from:
\begin{equation}
\sigma_{\rm Sample} = \sqrt{\frac{\mathop{\sum}\limits_{i}^{\rm Npix} \left( M_{{\rm ortho}, i} - M_{{\rm nom}, i} \right)^{2}}{\mathop{\sum}\limits_{i}^{\rm Npix} \left(\frac{1}{C_{{\rm ortho},i}} +\frac{1}{C_{{\rm nom},i}}\right)}}
\end{equation}
in which $M_{{\rm ortho}, i}$ and $M_{{\rm nom},i}$ are the flux densities in the $i$th pixel in the two maps out of 
which the jackknife is made (the flux in the jackknife map is $M_{{\rm ortho}, i} - M_{{\rm nom},i}$)
and $C_{{\rm ortho},i}$ and $C_{{\rm nom},i}$ are the numbers of timeline samples contributing to the $i$th 
pixel in the two maps. We measured the uncertainty on $\sigma_{\rm Sample}$ by randomly assigning 
the pixels to four groups and calculating $\sigma_{\rm Sample}$ separately for each group. We repeated this five times, which
was enough to give a reasonable estimate of the uncertainty in $\sigma_{\rm Sample}$.

We measured the noise per bolometer sample separately for the SGP and NGP fields and the values are given in 
Table~\ref{tab:SPIREinstrumental}. There is generally good agreement between the  two fields: a difference of
$\simeq$0.1 mJy at 250 and 350\micron\ and $\simeq$0.6 mJy at 500\micron. These differences are much greater
than the measured uncertainty; we do not know the reason for this but it does not 
seem likely to have any practical consequences. The averages of the noise values for the two fields are
31.38, 32.08, and 36.21\,mJy\,beam$^{-1}$ at 250, 350, and 500\micron, respectively. 
After adjusting for the small difference in average calibration (Section~\ref{sec:SPIREcal}),
these values are higher at 250 and 350\micron\ 
than those reported in \citetalias{Valiante2016} for the GAMA fields and lower at 500\micron,
but the difference is for all bands is $<$2\%.

\begin{figure*}
  \centering
  \includegraphics[trim=15mm 4mm 0mm 0mm,clip=True, width=0.7\textwidth]{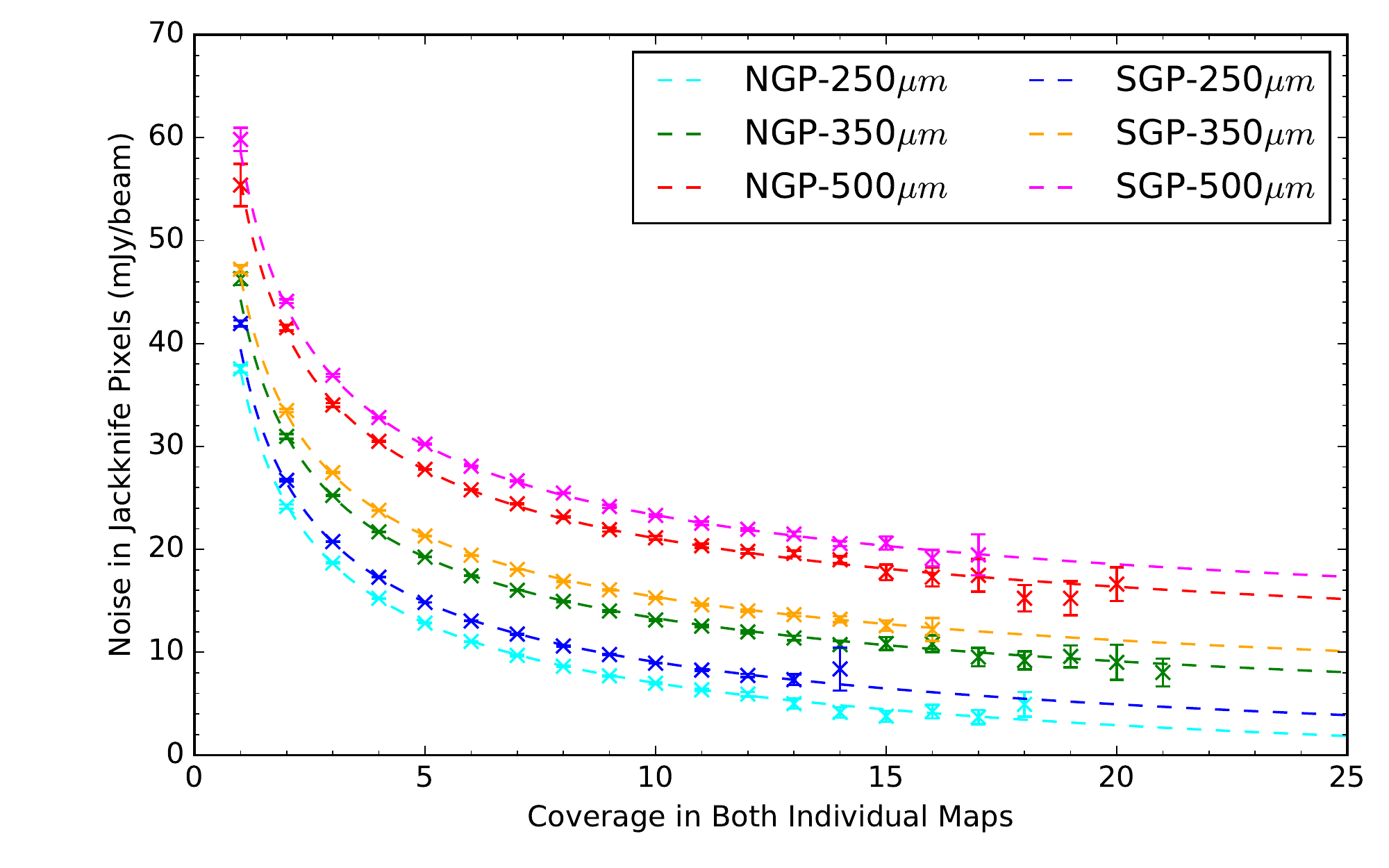}
  \figcaption{Data points show the standard deviation of the pixels in the jackknife map plotted
           against the number of timeline samples ($C_i$) in each pixel; we have only included pixels in which
           there were the same number of timeline samples in both maps used to make the jackknife.
           We only plotted values of $C_i$ for which there were at least 30 pixels in the jackknife map.
           To ensure the data points do not overlap,
           an offset of -7, -5, -1, +1, +5, +7 mJy\,beam$^{-1}$ has been applied to the 
           results for NGP-250\micron, SGP-250\micron,
           NGP-350\micron, SGP-350\micron, NGP-500\micron, SGP-500\micron, respectively. The dashed lines
           show the predicted noise from our noise-per-sample measurements assuming Gaussian noise.}
  \label{fig:jackknife}
\end{figure*}

A common assumption is that the instrumental noise in {\it Herschel}
images is Gaussian. We have tested this in Figure~\ref{fig:jackknife}.
We have divided the pixels in the jackknife maps by the number of timeline samples each
contains, only keeping pixels that have the same number of samples
in both of the maps used to make the jackknife. The figure shows
the noise per pixel plotted against
the number of timeline samples contributing to the pixel ($C_{i}$).
If the noise is Gaussian, we would expect the noise per pixel
to decrease as $C_{i}^{-1/2}$. The dashed lines in the figure
show the predictions of this Gaussian model, using our noise-per-sample estimates. 
The model agrees almost exactly with the observations, confirming that the
instrumental noise does have Gaussian statistical properties.

As part of the data release, we have produced maps showing the noise per pixel in the two fields. As our estimate of the
instrumental noise in each pixel, we have used $\sigma_{\rm Sample} / \sqrt{C_i}$,
in which $\sigma_{\rm Sample}$ is given in the top panel of Table~\ref{tab:SPIREinstrumental} and
$C_i$ is the number of timeline samples in each pixel. As a useful guide to the instrumental noise
in parts of the raw and matched-filtered images made with different numbers of scans, 
in Table~\ref{tab:SPIREinstrumental} (bottom panel) we
have given the average instrumental noise for map pixels produced from data from $\rm N_{scan}$ individual observations,
with values of $\rm N_{scan}$ from 1 to 7.

\begin{deluxetable*}{cccccccc}
\tablecaption{SPIRE Instrumental Noise \label{tab:SPIREinstrumental}}
\tablecolumns{8}
\tablewidth{0pt}
\tablehead{ & & \multicolumn{3}{c}{Raw Maps} & \multicolumn{3}{c}{Matched-Filtered Maps} \\
            & \multirow{2}{*}{Field} & \colhead{250\micron} & \colhead{350\micron} & \colhead{500\micron} & 
                                       \colhead{250\micron} & \colhead{350\micron} & \colhead{500\micron} \\
            & & \colhead{(mJy\,beam$^{-1}$)} & \colhead{(mJy\,beam$^{-1}$)} & \colhead{(mJy\,beam$^{-1}$)} & \colhead{(mJy\,beam$^{-1}$)} 
              & \colhead{(mJy\,beam$^{-1}$)} & \colhead{(mJy\,beam$^{-1}$)}
}
\startdata
Noise per & NGP & $31.327 \pm 0.005$ & $32.001 \pm 0.006$ & $35.922 \pm 0.012$ & - & - & - \\
Sample    & SGP & $31.426 \pm 0.005$ & $32.149 \pm 0.006$ & $36.506 \pm 0.006$ & - & - & - \\
\hline
\multirow{2}{*}{$\rm N_{scan} = 1$} & NGP &  $19.3715 \pm 0.0031$ &  $19.2201 \pm 0.0036$ &  $19.9354 \pm 0.0066$ & $10.1834 \pm 0.0018$ & $10.0136 \pm 0.0021$ & $15.8384 \pm 0.0060$ \\
                                    & SGP &  $17.9236 \pm 0.0030$ &  $17.8210 \pm 0.0032$ &  $19.4017 \pm 0.0034$ & $18.2855 \pm 0.0029$ & $18.5682 \pm 0.0031$ & $35.3218 \pm 0.0093$ \\
\multirow{2}{*}{$\rm N_{scan} = 2$} & NGP &  $10.3446 \pm 0.0017$ &  $10.1512 \pm 0.0019$ &  $11.1793 \pm 0.0037$ & $5.0488 \pm 0.0008$ & $4.9319 \pm 0.0010$ & $5.6003 \pm 0.0021$ \\
                                    & SGP &  $10.7022 \pm 0.0018$ &  $10.5098 \pm 0.0019$ &  $11.7208 \pm 0.0020$ & $5.2081 \pm 0.0008$ & $5.1019 \pm 0.0008$ & $5.8404 \pm 0.0015$ \\
\multirow{2}{*}{$\rm N_{scan} = 3$} & NGP & \ $8.6163 \pm 0.0014$ & \ $8.4736 \pm 0.0016$ & \ $9.3358 \pm 0.0031$ & $4.2170 \pm 0.0007$ & $4.1447 \pm 0.0009$ & $4.7175 \pm 0.0018$ \\
                                    & SGP & \ $8.9101 \pm 0.0015$ & \ $8.7633 \pm 0.0016$ & \ $9.7512 \pm 0.0017$ & $4.3557 \pm 0.0007$ & $4.2836 \pm 0.0007$ & $4.9041 \pm 0.0013$ \\
\multirow{2}{*}{$\rm N_{scan} = 4$} & NGP & \ $7.4953 \pm 0.0012$ & \ $7.3730 \pm 0.0014$ & \ $8.1258 \pm 0.0027$ & $3.6817 \pm 0.0006$ & $3.6268 \pm 0.0007$ & $4.1126 \pm 0.0016$ \\
                                    & SGP & \ $7.7019 \pm 0.0013$ & \ $7.5780 \pm 0.0014$ & \ $8.4179 \pm 0.0015$ & $3.7694 \pm 0.0006$ & $3.7243 \pm 0.0006$ & $4.2594 \pm 0.0011$ \\
\multirow{2}{*}{$\rm N_{scan} = 5$} & NGP & \ $6.9083 \pm 0.0011$ & \ $6.8135 \pm 0.0013$ & \ $7.5298 \pm 0.0025$ & $3.4204 \pm 0.0006$ & $3.3723 \pm 0.0007$ & $3.8342 \pm 0.0015$ \\
                                    & SGP & \ $6.9362 \pm 0.0012$ & \ $6.8602 \pm 0.0012$ & \ $7.6284 \pm 0.0013$ & $3.3952 \pm 0.0005$ & $3.3831 \pm 0.0006$ & $3.8603 \pm 0.0010$ \\
\multirow{2}{*}{$\rm N_{scan} = 6$} & NGP & \ $6.1648 \pm 0.0010$ & \ $6.0691 \pm 0.0011$ & \ $6.6935 \pm 0.0022$ & $3.0391 \pm 0.0005$ & $2.9969 \pm 0.0006$ & $3.4182 \pm 0.0013$ \\
                                    & SGP & \ $6.4155 \pm 0.0011$ & \ $6.3283 \pm 0.0011$ & \ $7.0319 \pm 0.0012$ & $3.1369 \pm 0.0005$ & $3.1289 \pm 0.0005$ & $3.5077 \pm 0.0009$ \\
$\rm N_{scan} = 7$                  & NGP & \ $5.6386 \pm 0.0009$ & \ $5.5154 \pm 0.0010$ & \ $6.0978 \pm 0.0020$ & $2.7739 \pm 0.0005$ & $2.7573 \pm 0.0006$ & $3.2051 \pm 0.0012$ \\
\enddata
\tablecomments{The instrumental noise properties of the SPIRE maps. The top two rows show the instrumental noise per
bolometer sample in the two fields. The other rows correspond to the average instrumental noise per map pixel  
for pixels with the same number of scans ($\rm N_{scan}$) for the raw (left) and matched-filtered maps (right).}

\end{deluxetable*}

\subsection{The SPIRE Confusion Noise}
\label{sec:SPIREconf}

By far a more difficult quantity to define and measure is the confusion noise 
(as \citetalias{Valiante2016} says, ``confusion is confusing''), and different scientific objectives
require different methods for measuring it. Source confusion has several different effects on
observations. Two of the most important are: (a) confusion increases the difficulty of
detecting sources by increasing the overall noise on an image; (b) confusion increases the error
on the flux measurements. \citetalias{Valiante2016} used two different definitions of confusion noise:
one suitable for measuring the signal-to-noise with which sources are detected on an image and 
one suitable for estimating the errors in flux measurements. We have used the same two definitions
of confusion, but improved the methods for measuring them described in \citetalias{Valiante2016}.

\begin{figure*}
  \centering
  \includegraphics[trim=0mm 3mm 0mm 4mm,clip=True, width=0.99\textwidth]{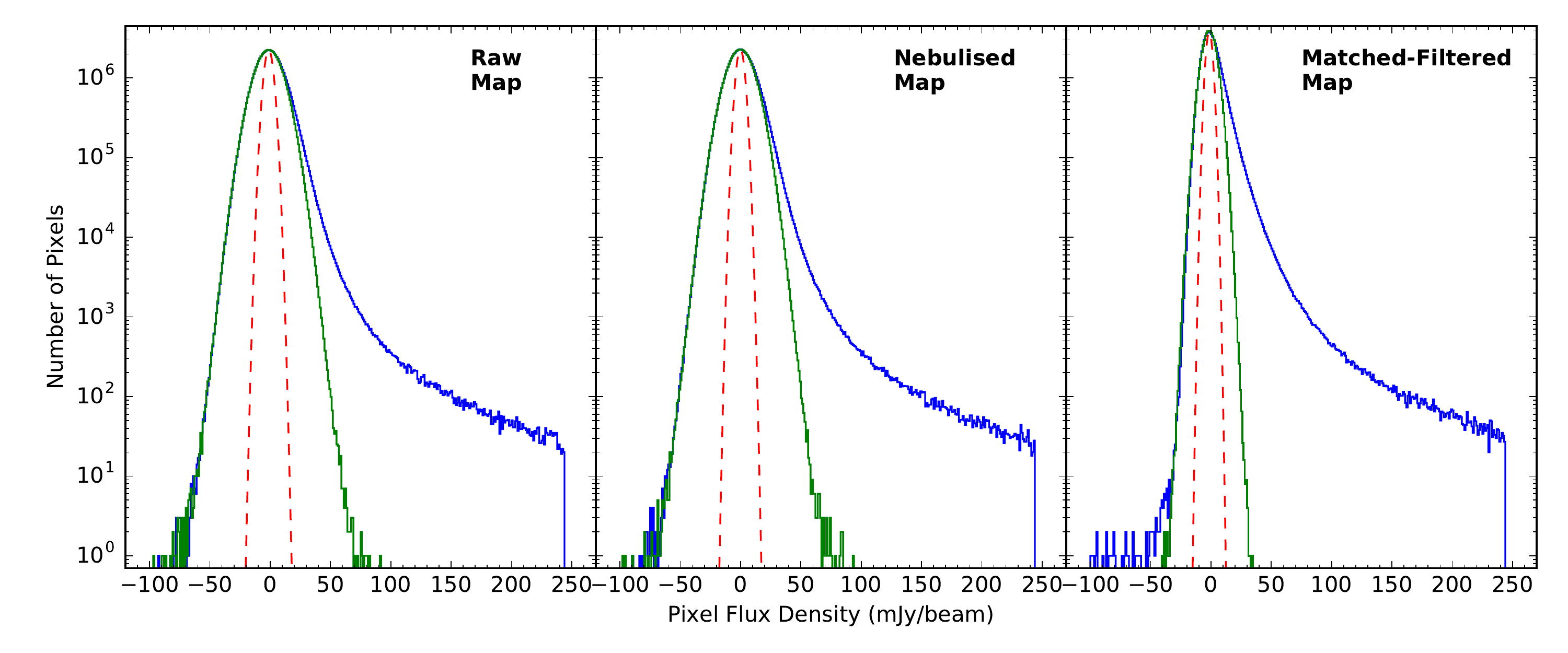}
  \figcaption{Histogram of pixel flux densities for the raw, nebulised and matched-filtered maps 
           of the NGP at 250\micron\ compared to
our model of the noise. The blue lines show the 
           real distribution of pixel fluxes on the map, the green is the histogram
           from our synthetic noise maps and the red dashed line shows the contribution just from confusion.}
  \label{fig:confusion1}
\end{figure*}

The first method was designed to produce a confusion estimate suitable for measuring the noise in signal-to-noise
estimates. \citetalias{Valiante2016} estimated this using the histogram of pixel values
in the SPIRE maps (Figure~\ref{fig:confusion1}). The shape of this distribution is produced by
the instrumental and confusion noise in the maps and the significant individual
sources, which produce the right-hand tail in the figure.
On the assumption that the tail of individual sources is not relevant for estimating the noise
in signal-to-noise estimates, \citetalias{Valiante2016} measured the confusion noise by first fitting a Gaussian to the
negative part of the pixel histogram, thus avoiding the positive tail,
and then calculating $\sqrt{\sigma_{\rm tot}^2 - \sigma_{\rm inst}^2}$, in which
$\sigma_{\rm tot}$ is the standard deviation of the best-fitting
Gaussian and $\sigma_{\rm inst}$ is the standard deviation of the Gaussian distribution for the instrumental noise.

Here we have used the same definition of confusion noise but a slightly different approach
for measuring it. We used the coverage map and noise-per-sample measurements
to generate an artificial image containing only instrumental noise. 
We then fit for the confusion noise (using the Python lmfit package) by
adding the confusion noise to our artificial map assuming the confusion noise has a Gaussian distribution.
A $\chi^2$ statistic is calculated from the difference between
the real pixel histogram and that from the model image;
the iteration with the lowest $\chi^2$ value gives our best estimate
of the confusion level. 
To avoid the biasing effect of the positive
tail produced by the significant sources, we generally only calculated $\chi^2$ for 
the negative side of the pixel histogram, with the bins
to the right of the peak contributing only if the value for the model lay
above the real distribution.
We repeated this whole process 96 times to produce
an estimate of the error on our measurement of the confusion
noise\footnote{Each time we used a different value of the noise per sample, generated from the
errors in Table 1, to allow for the uncertainty in this measurement.}.
The biggest advantage of this method over that
of \citetalias{Valiante2016} is that we are not assuming a single instrumental noise for the whole map.

Figure~\ref{fig:confusion1} shows the artificial histograms that produce the best fit to the histograms
for the raw 250\micron\ image of the NGP, the 250\micron\ image from which the background has been subtracted using
{\it Nebulizer}, and the 250\micron\ image from which the background has been subtracted
and which has then been convolved with the matched filter. The values of the confusion noise that
give the best fit to the data for both the SGP and NGP 
are given in Table~\ref{tab:SPIREconfusion-hist} for all three wavebands.
Our confusion noise values are the most accurate values produced at the SPIRE wavelengths, with both
fields agreeing to within 0.1\,mJy\,beam$^{-1}$ for the nebulised and matched-filtered maps in all bands (the
exception
is SPIRE 250\micron\ where the difference is 0.14\,mJy/beam). This is a significant improvement over \citetalias{Valiante2016}
whose estimates varied by up to 0.9\,mJy/beam between fields. 
Our confusion values tend to be slightly lower than those of V16 on the nebulised 
images and slightly higher on the matched-filtered images, which is probably due
to the improved method we have used in this paper.

\begin{deluxetable*}{ccccc}
\tablecaption{SPIRE Confusion Noise \label{tab:SPIREconfusion-hist}}
\tablecolumns{5}
\tablewidth{0pt}
\tablehead{
     &\multirow{2}{*}{Field} & \colhead{250\micron} & \colhead{350\micron} & \colhead{500\micron} \\
     &   & \colhead{(mJy\,beam$^{-1}$)} & \colhead{(mJy\,beam$^{-1}$)} & \colhead{(mJy\,beam$^{-1}$)}
}

\startdata
\multirow{2}{*}{Raw Map} & NGP & $3.366 \pm 0.004$ & $4.517 \pm 0.021$ & $4.555 \pm 0.011$ \\
                         & SGP & $3.516 \pm 0.010$ & $4.567 \pm 0.031$ & $5.503 \pm 0.012$ \\
Nebulised & NGP & $3.194 \pm 0.017$ & $4.129 \pm 0.041$ & $4.414 \pm 0.018$ \\
Map       & SGP & $3.050 \pm 0.015$ & $4.138 \pm 0.016$ & $4.495 \pm 0.039$ \\
Matched-Filter & NGP & $2.483 \pm 0.017$ & $3.257 \pm 0.005$ & $4.436 \pm 0.015$ \\
Map            & SGP & $2.470 \pm 0.045$ & $3.249 \pm 0.005$ & $4.490 \pm 0.018$ \\
\enddata
\tablecomments{The confusion noise estimated by fitting using the histogram fitting method described in Section~\ref{sec:SPIREconf}.}
\end{deluxetable*}

Although the confusion estimates for the NGP and SGP agree well, our uncertainty estimates
are so small that the differences between the NGP and SGP are formally
significant.
Although these fields are very large, it is possible that these differences are due to
large-scale interstellar cirrus or to large-scale extragalactic  structure \citep{Negrello2017}.
Part of the explanation may be that the instrumental
noise is such a large part of the total noise, especially in the raw
maps, that small errors in
the estimate of instrumental noise
may lead to large errors in the estimate of confusion noise.
Another possible problem may be that our assumption that the source
population can be divided into a population of faint confusing
sources, which produces a Gaussian pixel distribution, and a population
of sources which are detected individually and which produce a non-Gaussian
tail to the pixel distribution, may be too simplistic.

The second definition of confusion noise used by \citetalias{Valiante2016} was one designed to produce
an estimate suitable for estimating the errors on flux measurements.
Errors on flux measurements are produced by all the other sources on the image, not just
the faint ones contributing to the Gaussian distribution in Figure~\ref{fig:confusion1} 
but also the tail of significant sources. The only sources on an image that
cannot contribute to the flux error for a source are the pixels in the map that are brighter than that source.
To produce an estimate of the confusion noise appropriate for a source with flux density $F_s$, 
\citetalias{Valiante2016} measured the variance on an image but only included pixels with flux densities $<F_s$.
They then took the confusion noise as  $\sqrt{\sigma_{\rm var}^2 - \sigma_{\rm inst}^2}$,
in which $\sigma_{\rm var}^2$ is the variance and $\sigma_{\rm inst}$ is the standard deviation
of the Gaussian distribution for the instrumental noise (see \citetalias{Valiante2016} for additional details).

\begin{figure*}
  \centering
  \includegraphics[trim=12mm 4mm 5mm 0mm,clip=True, width=0.99\textwidth]{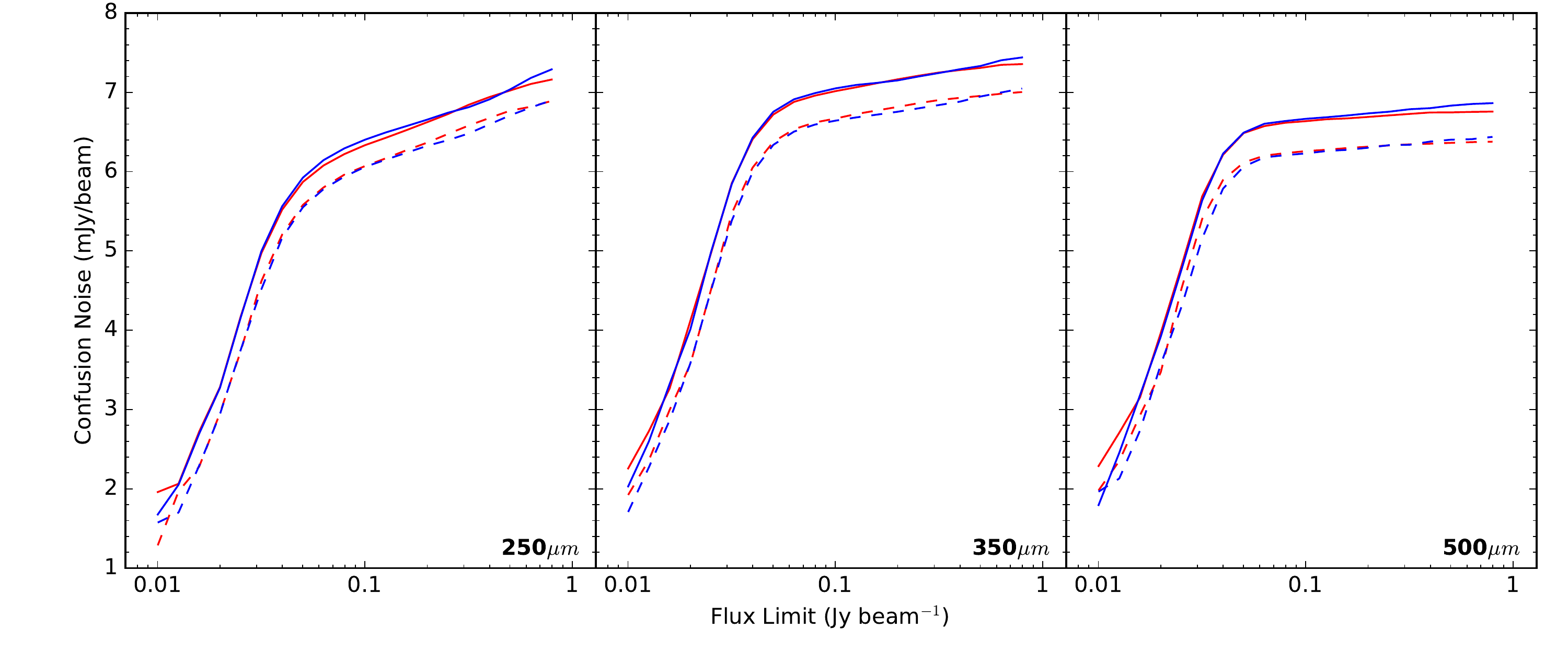}
  \figcaption{Estimates of confusion noise in the the three SPIRE wavebands using 
           our second method. The red and blue coloured lines
           represent measurements of the NGP and SGP, respectively. The solid lines are the measurements
           on the raw-SPIRE maps, while the dashed lines are measurements on the nebulised maps. 
           Uncertainties are not shown as they are too small to plot. The values used to make this plot are
           given in Table~\ref{tab:confFlim} in Appendix~\ref{app:confusion}.}
  \label{fig:confusion2}
\end{figure*}

The only difference between our method and that of \citetalias{Valiante2016} is that we have taken account of
the variation in depth over the image arising from different coverage levels, $C_i$.
We have allowed for this by estimating the confusion noise
separately for pixels with different numbers of bolometer samples and then averaging the different
estimates of the confusion noise. 

Figure~\ref{fig:confusion2} shows the results for the raw maps and for the maps from which the
background has been subtracted using {\it Nebuliser}. As expected, the confusion noise for the images from which
the background (mostly cirrus emission) has not been subtracted is higher than for the
images from which the background has been subtracted. The results for the NGP and SGP are almost the
same. By definition, the confusion noise depends on $F_s$. As in \citetalias{Valiante2016}, we
use a value of $F_s$ of 200 mJy to estimate a confusion noise that is easy to compare with the
estimates of others. For the background-subtracted maps and with $F_s=200$ mJy,
we find the values of the confusion noise are 6.62, 7.16, 6.69\,mJy\,beam$^{-1}$ at 250, 350, and 500\micron, respectively, 
for the NGP field, and 6.66, 7.15, 6.73\,mJy\,beam$^{-1}$ at 250, 350, and 500\micron, respectively, for the SGP field.
\citetalias{Valiante2016} found mean values of 6.53, 7.03, and 6.58\,mJy\,beam$^{-1}$ for the three GAMA fields at 250, 350,
and 500\micron, respectively (corrected for the calibration differences), using a very similar method. 
As for the 
previous method, the differences between the NGP and
the SGP are much smaller than those between the GAMA fields
found by \citetalias{Valiante2016}.
\citet{Nguyen2010} estimated the confusion noise
in the HERMES survey, using a fairly similar technique to this second method, and derived 
estimates of 5.8, 6.3, and 6.8\,mJy\,beam$^{-1}$\footnote{We have not corrected for changes in flux calibration but these
changes are much smaller than the
differences in the confusion estimates.}.
The estimates in our work using this technique are broadly similar to those of \citetalias{Valiante2016} but systematically 
higher than those of \citet{Nguyen2010}. It is unknown whether
these differences are genuinely the result of differences in the source populations in the
different fields, or due to the differences in the 
methods used in the different fields.

By combining our histogram confusion estimates with those of the instrumental noise from Section~\ref{sec:SPIREinst}, 
we find the total pixel noise in our nebulised images of 11.0, 11.1, and 12.3\,mJy\,beam$^{-1}$ at 250, 350, and 500\micron, respectively. 
If the matched filter is used to extract point sources is applied to the map, our 1$\sigma$ map sensitivity estimate 
is 5.7, 6.0, and 7.3\,mJy\,beam$^{-1}$, respectively.

\subsection{Photometry for Point and Extended Sources}
\label{sec:spirePhot}

The correct method to use for photometry depends on whether the object is expected to
be extended or unresolved by the SPIRE PSF. If the object is unresolved, the
best method is to use the flux value at the object's position
on the SPIRE image that has been convolved with the matched filter.
As part of the data release, we have produced a map of the instrumental
noise on this image. However,
the error on the flux density will also include a component from the confusion noise.
The correct value to use for the confusion noise depends on the
purpose of the investigation (see the previous section), but for detection
experiments we suggest using the values
obtained from fitting the Gaussian part of the 
pixel histograms, which are given in Table~\ref{tab:SPIREconfusion-hist}.
For an estimate of the error of the flux density
of an individual source, the correct
confusion noise value to use would be that obtained from measuring
the variance on the image, which depends on the flux density
of the source, as shown in Figure~\ref{fig:confusion2}, and given in 
Table~\ref{tab:confFlim} of Appendix~\ref{app:confusion}.
Whichever version of the confusion noise is chosen, the confusion noise and
the instrumental noise, taken from the map of instrumental noise,
should be added in quadrature.

Astronomers interested in 
carrying out a statistical ``stacking analysis'', in order
to measure the
mean submillimeter flux density of
some class of object, 
should use the 
images that have been nebulised to remove cirrus emission.
They should be aware that the means of the maps are not necessarily zero and so
they should subtract the mean from the map before carrying out
the analysis\footnote{{\it Nebuliser} produces the best estimate
of the sky value at each position but this value is
not generally equal to the mean in that region.
Therefore, stacking analyses, which sum the emission from large numbers
of sources will be sensitive to any small systematic error
in the way {\it Nebuliser} estimates the sky value. To be on the safe
side, we therefore recommend that a stacking analysis should only
be carried out after the mean has been subtracted from the image.}.
We recommend using the SIMSTACK algorithm \citep{Viero2013}
or similar, that allows the user to correct for the effects of clustering.
We would recommend that astronomers interested in carrying out a
stacking analysis should also carry out a Monte Carlo simulation
in which they measure the mean flux density at a large number of random
positions. This procedure will (a) give an estimate of the mean level on
the map and (b) produce an empirical estimate of the error in the stacking
measurements.

Photometry of extended sources should be carried out using
aperture photometry. The images supplied in 
the data release have units of Jy\,beam$^{-1}$. These can be converted into
images with units of Jy\,pixel$^{-1}$, suitable for aperture photometry,
by dividing the flux value in each pixel by a factor
$C_{\rm conv}$, which is given by the area of the telescope's beam divided
by the area of a pixel. The current values in the
SPIRE Data Reduction Guide\footnote{\url{http://herschel.esac.esa.int/hcss-doc-15.0/index.jsp\#spire\_drg}} 
are 469.4, 831.2 and 1804.3\,arcsec$^2$ at 250, 350, and 500\micron, respectively.
Note that it is possible to produce SPIRE maps that have been optimized
for aperture photometry, using the SPIRE `relative gain' method. However,
for simplicity, we decided to produce only a single set of maps, optimized
for point-source detection.

We recommend carrying out aperture photometry on
the images from which the background has been subtracted with {\it Nebuliser}.
The application of {\it Nebuliser} does mean that the flux density of any sources with a size greater than 3\arcmin\
might be underestimated,but our tests on the GAMA fields found no 
evidence of this effect \citepalias{Valiante2016}. 
We also found that the photometric errors were smaller
if we used the {\it nebulised} images.
Note that in this case there is no need to subtract
the mean map value, since the application of {\it Nebuliser} should already
have subtracted the best estimate of the sky level at that position.
Some of the object's emission will fall outside the aperture because of the extended profile of the PSF 
\citep{Griffin2013}. As part of the data release, we have supplied a table of corrections factors for this effect.

\begin{figure}
  \centering
  \includegraphics[trim=20mm 12mm 4mm 4mm,clip=True, width=0.49\textwidth]{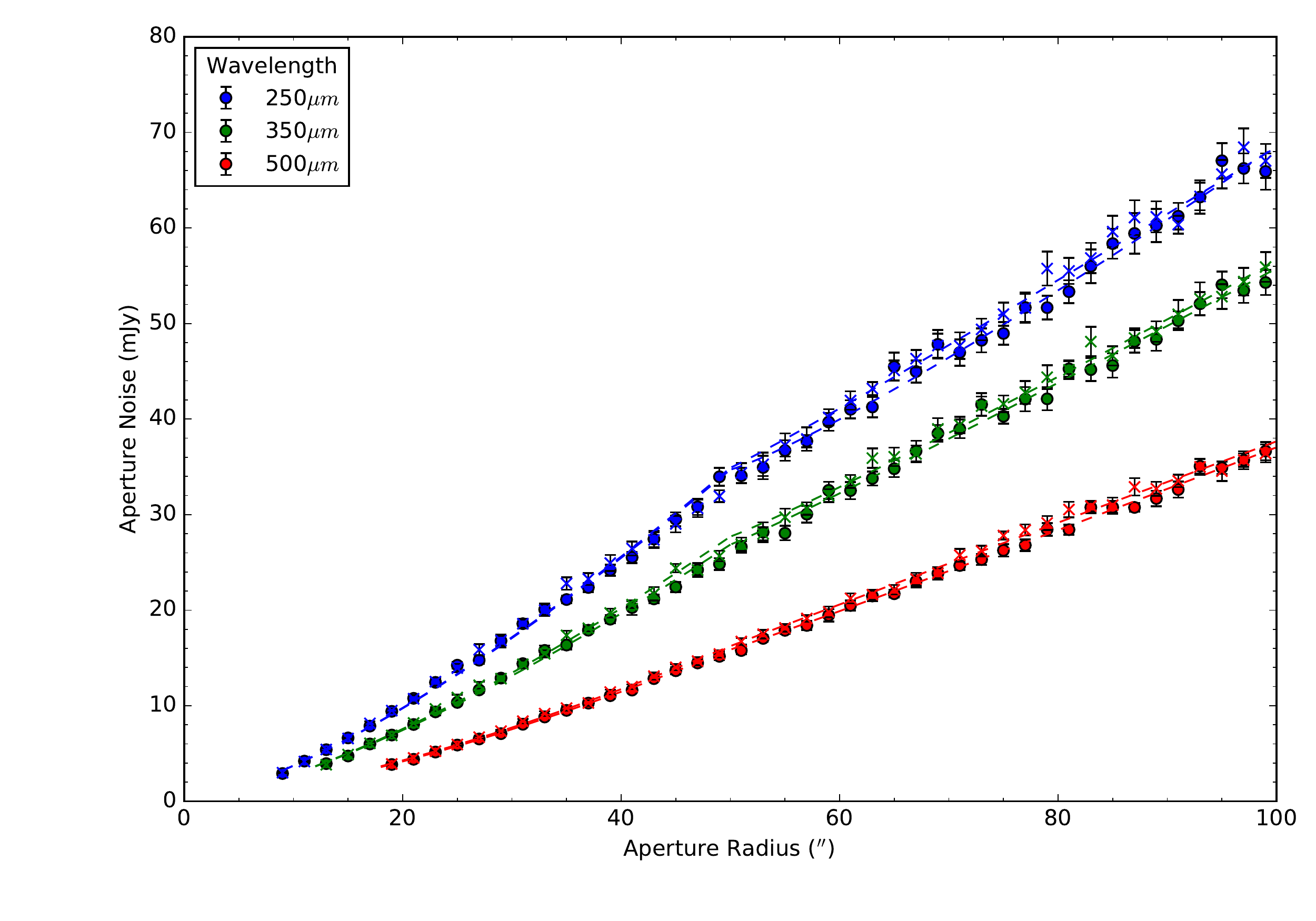}
  \figcaption{The results of the Monte Carlo simulation 
of the flux-density errors for aperture photometry on the SPIRE images
(see the text for details). The figure shows our estimate of the flux density
error plotted against the radius of the aperture.
           Results from the NGP and SGP are shown by the circular and cross points, 
           respectively. The best-fit power-law
models described in Section~\ref{sec:spirePhot} are shown by the dashed lines.}
  \label{fig:monte}
\end{figure}

We have carried out Monte Carlo simulations to estimate the errors in 
the flux densities measured with aperture photometry. We placed apertures
randomly on the NGP and SGP maps in areas which are made from two individual 
observations ($\rm N_{scan}=2$). The aperture radii were
varied in size from approximately the beam size up to 100\arcsec, in
2\arcsec\ intervals, and for each radius we used 3000 random
positions. Figure~\ref{fig:monte} shows the 
results of the Monte Carlo simulation, and very consistent results between the two fields can be seen.

We assumed the relationship between flux error and radius is a
power law, since if the noise is dominated by instrumental random noise we 
should get a simple linear relationship: flux error
$\propto$ radius. We found that we needed to use two power laws to 
fully describe the relationship at all radii, with the change in relationship occurring at 50\arcsec. 
Our model is described by:
\begin{equation}
 \sigma_{\rm ap} ({\rm mJy}) = \begin{cases}
    A r^\alpha & \text{if $r \leq 50^{\prime\prime}$}\\
    B \left(r-50\right) ^\beta + A 50^{\alpha} & \text{for $r > 50^{\prime\prime}$}
 \end{cases}
 \label{equ:apModels}
\end{equation}
where $\sigma_{\rm ap}$ is the flux error in mJy and $r$ is the radius in arcseconds. The best-fit values for this
relationship for all bands and fields are given in Table~\ref{tab:SPIREaperture}. Above 
a radius of 50\arcsec\ the relationship
is quite similar to that expected for pure instrumental
noise,
with values of $\beta$ between 0.98 and 1.17.
Below 50\arcsec\ the relationship is much steeper with values of $\alpha$
between 1.37 and 1.48. 
This may be due to small-scale cirrus emission, which would not have been
removed with the filtering scale used in {\it Nebuliser}, or possibly
some effect of source confusion.

As the areas where $\rm N_{scan} > 2$ are limited in size,
we are unable to perform a Monte Carlo simulation for the deeper regions. As we believe we understand the 
properties of the instrumental
noise, we can account for the differences in $\rm N_{scan}$ by subtracting our instrumental noise for $\rm N_{scan} = 2$
in quadrature and adding back in quadrature the appropriate noise (as measured in Section~\ref{sec:SPIREinst}, and
tabulated in Table~\ref{tab:SPIREinstrumental}). The relationship between the
flux error and aperture radius, for any value of $\rm N_{scan}$, is then given by:
\begin{equation}
 \scriptsize
 \sigma_{\rm ap} ({\rm mJy}) = \begin{cases}
    \sqrt{\left(A r^\alpha\right)^2 - X \left(\sigma_{\rm inst,2}^2 - \sigma_{\rm inst,N}^2\right)r^2} \\
    \hfil \hfil \text{if $r \leq 50^{\prime\prime}$}\\
    \sqrt{\left(B \left(r-50\right) ^\beta + A 50^{\alpha}\right)^2- X \left(\sigma_{\rm inst,2}^2 - \sigma_{\rm inst,N}^2\right)r^2}\\
    \hfil \hfil \text{for $r > 50^{\prime\prime}$}
 \end{cases}
 \label{equ:apModels-Ncorrect}
\end{equation}
where $A, B, \alpha, \beta$ are the same as in Equation~\ref{equ:apModels}, $X$ is a constant given in Table~\ref{tab:SPIREaperture}
(which varies between bands) to account for beam area and pixel size, 
and $\sigma_{\rm inst,N}$ (mJy\,beam$^{-1}$) is the instrumental noise for $\rm N_{scan} = N$ as given in 
Table~\ref{tab:SPIREinstrumental}. 
This equation uses the values of instrumental noise averaged over all the pixels
with the same number of scans (Table~\ref{tab:SPIREinstrumental}). Purists interested in using the
actual instrumental noise at the position of a source can measure this
instrumental noise from the noise map, and then obtain the total flux
error by modifying equation~\ref{equ:apModels-Ncorrect} in a fairly obvious way. If users wish to use
elliptical apertures they could either run their own Monte Carlo simulation on the released maps, 
or a reasonable estimate of the flux error can be obtained by using the 
estimate for a circular aperture with the same area.

The width of the SPIRE filters mean that both the size of the
PSF and the power detected by SPIRE depend on the spectral energy distribution (SED) of the
source. The SPIRE data-reduction pipeline is based on the assumption
that the flux density of a source depends on frequency$^{-1}$, and
all our images are ultimately based on this assumption. If the user has
reason to know the SED of a source, the flux densities
should be corrected
using the corrections from either Table~5.7
or 5.8 from the SPIRE handbook\footnote{\url{http://herschel.esac.esa.int/Docs/SPIRE/spire_handbook.pdf}}. 
It is important to apply these corrections, since they can be quite large: for
a point source with a typical dust spectrum ($T$=20\,K, $\beta$=2) the multiplicative 
correction is 0.96, 0.94, and 0.90 at 250, 350, and 500\micron, respectively. 

Finally, on top of the other flux density errors, there is
an error from the uncertainty in the basic flux calibration of the instrument.
At the time of writing, the error in the flux density
arising from the uncertainty in the absolute flux density
of Neptune is 4\%, and there is an additional 1.5\% error
that is uncorrelated between the SPIRE bands (SPIRE Data Reduction Guide).
The current recommendation is that these factors should be added linearly, and 
so the reader should use
a calibration error of 5.5\%.

\begin{deluxetable}{ccccccc}
\tablecaption{Aperture Noise Model Best-fit Parameters \label{tab:SPIREaperture}}
\tablecolumns{7}
\tablewidth{0pt}

\tablehead{\colhead{Waveband} & \colhead{Field} & \colhead{$A$} & \colhead{$\alpha$} & \colhead{$B$} & \colhead{$\beta$} & \colhead{$X$}}

\startdata
\multirow{2}{*}{100\micron} & NGP & 0.749 & 1.475 & 6.244 & 0.971 & \multirow{2}{*}{--} \\
                            & SGP & 0.720 & 1.473 & 6.235 & 0.970 & \\
\multirow{2}{*}{160\micron} & NGP & 0.642 & 1.444 & 4.193 & 0.995 & \multirow{2}{*}{--} \\
                            & SGP & 0.620 & 1.446 & 4.247 & 0.992 & \\
\multirow{2}{*}{250\micron} & NGP & 0.152 & 1.388 & 0.336 & 1.179 & \multirow{2}{*}{5.13$\times 10^{-4}$} \\
                            & SGP & 0.164 & 1.368 & 0.527 & 1.066 & \\
\multirow{2}{*}{350\micron} & NGP & 0.117 & 1.389 & 0.539 & 1.016 & \multirow{2}{*}{2.91$\times 10^{-4}$} \\
                            & SGP & 0.111 & 1.410 & 0.497 & 1.016 & \\
\multirow{2}{*}{500\micron} & NGP & 0.052 & 1.464 & 0.372 & 1.033 & \multirow{2}{*}{1.39$\times 10^{-4}$} \\
                            & SGP & 0.056 & 1.451 & 0.459 & 0.984 & 
\enddata
\tablecomments{The best-fit parameters for the relationship between flux error and aperture radius
(Equation~\ref{equ:apModels}). See Section~\ref{sec:spirePhot} for details. The $X$ column gives the 
constant required to correct the relationship for regions of the map with different $\rm N_{scan}$ values
(see Equation~\ref{equ:apModels-Ncorrect}).}

\end{deluxetable}

\subsection{Power Spectrum of SPIRE Maps}
\label{sec:powerSpec}

\begin{figure}
  \centering
  \includegraphics[trim=14mm 2mm 0mm 2mm,clip=True, width=0.49\textwidth]{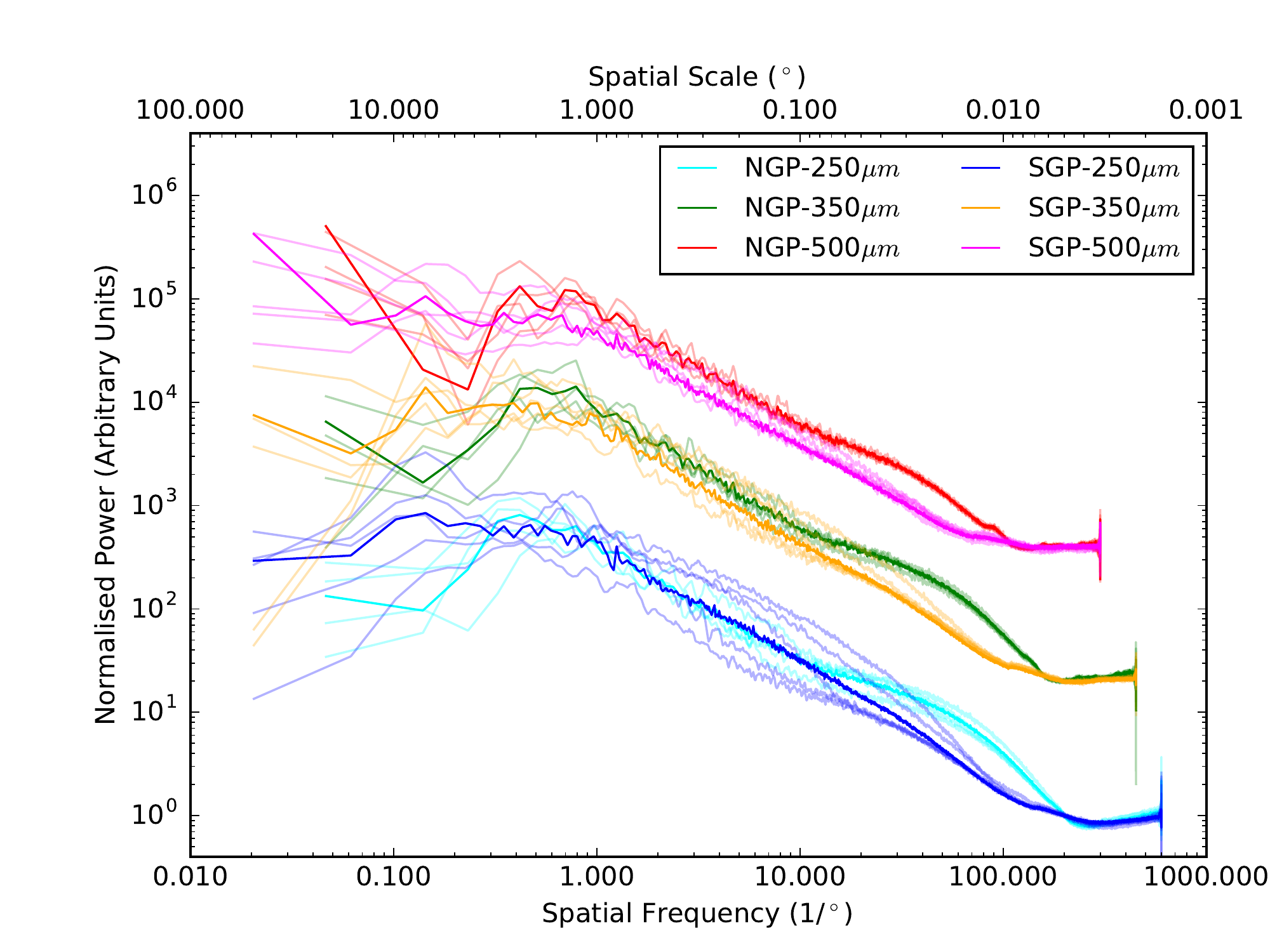}
  \figcaption{The 1D angular power spectrum of our raw maps for the NGP and SGP fields in the three SPIRE
                      bands. The dark lines show the power spectrum for the entire mosaic, while the lighter lines
                      are for the individual tiles in each mosaic. The profiles in each band are normalized so they
                      have the same value at 0.005$^{\circ}$.}
  \label{fig:allPowerSpec}
\end{figure}

The primary science goals of \hatlas\ are to investigate individual sources, and so our maps were made to
optimize the detection and flux-extraction of these small-scale structures. 
\citet{Pascale2011} used simulations of our observing strategy and map-making techniques 
to show that there is attenuation of the structure in the \hatlas\ maps on scales $>$20\arcmin. 
Since \citet{Pascale2011} all-sky maps produced by the \textit{Planck} observatory 
\citep{Planck2011} have been released \citep{Planck2016} which provide a useful
`truth' map to compare with the \hatlas\ maps at 350\micron\ and 500\micron. In this section, we calculate 
the 1D angular power spectrum of our maps (using the agpy package\footnote{https://github.com/keflavich/agpy/})
to investigate what emission scales are preserved in our maps, and if our maps are consistent between tiles.

In Figure~\ref{fig:allPowerSpec} we show the 1D power spectrum from the raw SPIRE maps, as well as for each
individual tile in the mosaics. The power spectra for each field tend to be in good agreement with each other,
especially at 350 and 500\micron. The differences between fields is most likely explained due to 
differences in the cirrus emission. Given the good agreement between tiles in a field, 
an average `transfer function' describing the depression of power as a function of angular scale,
could be used for each of the two fields.

\begin{figure}
  \centering
  \includegraphics[trim=14mm 2mm 0mm 2mm,clip=True, width=0.49\textwidth]{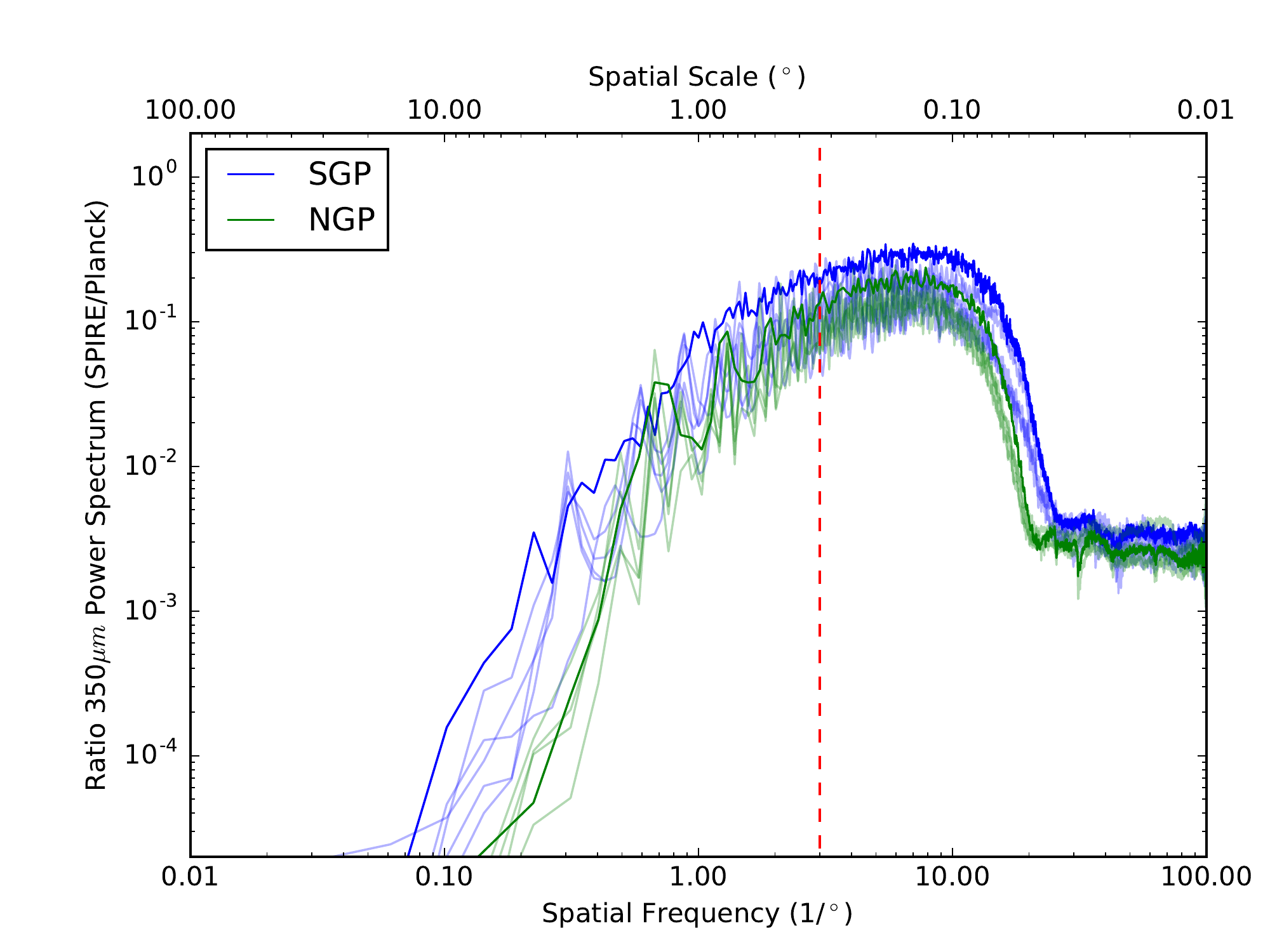}
  \figcaption{Ratio of the 1D angular power spectra (SPIRE/\textit{Planck}) for the NGP 
                      and SGP fields at 350\micron. As for Figure~\ref{fig:allPowerSpec} the dark lines are
                      for the entire field and the lighter lines are for individual tiles. The 
                      red dashed line shows the 20\arcmin\ scale found by \citet{Pascale2011}, where attenuation
                      of emission starts.}
  \label{fig:relPowerSpec}
\end{figure}

To test whether the differences seen in Figure~\ref{fig:allPowerSpec} are due to variations in the cirrus emission we
compare our maps at 350\micron\ with the \textit{Planck} Public Release 2 maps \citep{Planck2016}. We first convolved
both the NGP and SGP maps to the same resolution as \textit{Planck} 
using the effective \textit{Planck} beam for our field. Both the \textit{Planck} and 
SPIRE maps were matched to the same 36\arcsec\ pixel grid and converted to the same flux-density units. 
Figure~\ref{fig:relPowerSpec}
shows the ratio of the 1D power spectra of the SPIRE and \textit{Planck} maps. The ratio maps show broad agreement 
across all individual tiles and mosaics, confirming that the differences in Figure~\ref{fig:allPowerSpec} is due
to cirrus emission. At small spatial scales the 
low ratios of SPIRE/\textit{Planck} power are due to the greater sensitivity of the \hatlas\ observations. The low ratios at
higher spatial scale are due to the finite size of the maps and the 
fact that a single SPIRE observation made from scanning the telescope in a single direction will miss large-scale
power in the direction orthogonal to the scan direction \citep{Waskett2007}.
Our results show the attenuation of emission begins on a scale of $\sim$15\arcmin, broadly in agreement
with the value of 20\arcmin\ found by \citet{Pascale2011}. It is possible that our observations affected by `cooler burps'
(see Section~\ref{sec:SPIREtimelines}) could have greater attenuation, but, due to the complexity of isolating these regions we did not investigate this further.

In principle, it is possible to correct for this attenuation on large scales using an alternative map-maker 
\citep{Waskett2007}. If users wish to create maps with alternative map-makers,
the authors can be contacted to assist with data/customised timelines. For the 350 and 500\micron\ bands it is also possible
to combine the \Hersc\ and \textit{Planck} data to create maps that have the correct power on all scales. Of course,
the images on large scales will also be affected by cirrus. 
        
\section{Photometry on the PACS Maps}

\subsection{The PACS Instrumental and Confusion Noise}
\label{sec:pacs}

The PACS maps are very different from the SPIRE maps. The higher
instrumental noise means that source confusion is less important and
the instrumental noise is correlated between pixels. 
It is more challenging to measure the confusion and instrumental
noise on the PACS maps  because 
\textit{Jscanamorphos} 
uses the existence of multiple PACS observations to remove the effect
of temporal changes in the detectors, which means that it
is not possible to use
jackknifes to estimate the instrumental noise.

To estimate the PACS confusion noise, we used 
a similar approach to that of \citet{Magnelli2013}, who estimated the confusion
in the GOODS-S field. We measured the total noise in regions of the map with different number of observations
as seen in the $\rm N_{scan}$ maps shown in Figure~\ref{fig:pacsNGP} and \ref{fig:pacsSGP}. 
To measure the noise, we fitted a Gaussian to the
negative part of the histogram of pixel values, using the positive side
as an upper limit (similar to what we did for SPIRE in Section~\ref{sec:SPIREconf}). 
This gave us a plot of $\sigma_{\rm pix}$ versus $\rm N_{scan}$.
We then fitted a simple model to this relationship
The
model has an instrumental noise component, which scales with the
number of observations ($\rm N_{scan}$)
contributing to each pixel, and a constant confusion term:
\begin{equation}
  \sigma_{\rm pix} ({\rm mJy}) = \sqrt{\left(\sigma_{\rm inst}\ N_{\rm scan}^{-0.5}\right)^2 + \sigma_{\rm conf}^2}
  \label{equ:pacsNoiseModel}
\end{equation}
where $\sigma_{\rm pix}$ is the total pixel noise in mJy, $\sigma_{\rm inst}$ is the instrumental noise in mJy for a single PACS
observation ($\rm N_{scan}$=1) , and $\sigma_{\rm conf}$ is the confusion noise (in mJy).
In principle, this procedure allowed us to estimate
$\sigma_{\rm inst}$ and $\sigma_{\rm conf}$.

\begin{figure*}
  \centering
  \includegraphics[trim=0mm 2mm 0mm 0mm,clip=True, width=0.99\textwidth]{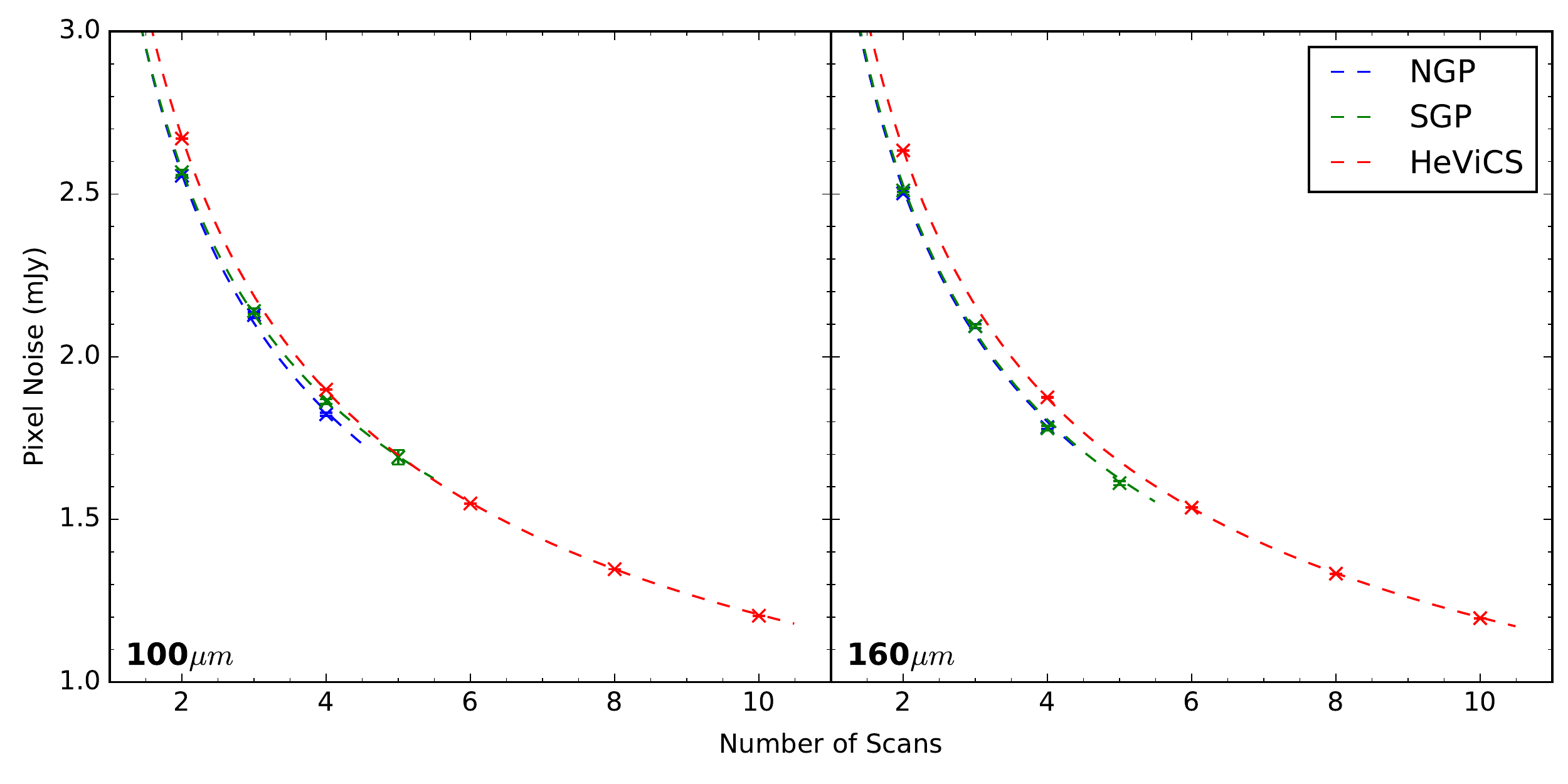}
  \figcaption{Pixel noise in PACS tiles versus
           number of observations contributing
to each pixel ($\rm N_{scan}$ - see Figure~\ref{fig:pacsNGP} and \ref{fig:pacsSGP}). 
The blue, green, and red crosses show the
           measurements from the NGP, SGP and \hevics\ fields, respectively.
           The dashed lines show the best fit of equation~\ref{equ:pacsNoiseModel} to each dataset.}
  \label{fig:PACSnoise}
\end{figure*}

We initially applied this method to our final maps,
but found that the noise in the regions in which the tiles overlap is significantly
reduced due to the re-projecting procedure used to create the mosaics\footnote{In overlapping areas
in which $\rm N_{scan} = 2$ the noise is reduced by a factor of 0.90, 0.91, 0.84 and 0.86
for the NGP 100\micron, NGP 160\micron, SGP 100\micron\ and SGP 160\micron, respectively.}.
We therefore decided to use only the individual tiles, which limited the range of
$\rm N_{scan}$ to 2--5, reducing the sensitivity of the method. To regain the sensitivity, we
used some observations from the \hevics\ survey \citep{Davies2010},
which mapped $\sim$55 sq. deg. of the Virgo Cluster using the same fast-scan parallel observing mode
that we used. While most of the Virgo Cluster
was observed in $4^\circ\times4^\circ$ tiles with eight observations per field, the
northernmost Virgo tile was observed 10 times by PACS. We reduced the observations of this 
tile using the same \textit{Jscanamorphos} method we used for \hatlas, starting
with the level-1 data produced by the standard pipeline.
We then made five independent maps from each pair of observations and
applied {\it Nebuliser} to each map, 
which gave us five maps of the same region of sky.
We then averaged 
various combinations of maps and estimated the total noise on each combined
map using the method above, giving us estimates
of the total noise from 
$\rm N_{scan} = $
2--10. The results of pixel noise versus $\rm N_{scan}$ are shown in Figure~\ref{fig:PACSnoise} 
for both the \hatlas\ and \hevics\ results.
The values of $\sigma_{\rm inst}$ and $\sigma_{\rm conf}$ obtained
from fitting
Equation~\ref{equ:pacsNoiseModel} to the results for the individual fields
are given in Table~\ref{tab:pacsPixNoise}.

As expected, the estimates of instrumental noise in Table~\ref{tab:pacsPixNoise} are much
higher than the estimates of confusion noise.
The errors on the confusion noise estimates
for the \hevics\ field
are much less than those for the NGP and SGP fields because of the
larger range of $\rm N_{scan}$. The confusion noise estimates for
the different fields are formally inconsistent, which we suspect arises because
the instrumental noise is so much larger than the confusion noise, making
any estimate for estimating the confusion noise sensitive to systematic
errors (e.g. if the assumption that the noise is Gaussian is slightly wrong).
Our most reliable estimates of confusion noise come from the \hevics\ tile
because of the larger range of $\rm N_{scan}$, and are
 $0.184 \pm 0.004$ and $0.240 \pm 0.003$\,mJy, at 100 and
160\micron, respectively. These estimates are broadly similar at 100\micron, but differ at 160\micron, to those
presented by \citet{Magnelli2013} of 0.15 and 0.68\,mJy at 100 and 160\micron, although these
values may not be directly comparable due to differences in pixel size and beam size. 
Assuming beam areas of 207 and 308\,arcsec$^2$ at 100 and 160\micron\ (calculated from our measured PSFs),
the confusion noise is 4.23 and 4.62\,mJy\,beam$^{-1}$ at 100 and 160\micron, respectively.

\begin{deluxetable}{cccc}
\tablecaption{PACS Pixel Noise Model Parameters \label{tab:pacsPixNoise}}
\tablecolumns{4}
\tablewidth{0pt}

\tablehead{\multirow{2}{*}{Waveband} & \multirow{2}{*}{Field} & \colhead{$\sigma_{\rm inst}$} & \colhead{$\sigma_{\rm conf}$} \\
                          &                        & \colhead{(mJy)} & \colhead{(mJy)}}

\startdata
\multirow{3}{*}{100\micron} & NGP & 3.578$\pm$0.013 & 0.393$\pm$0.047 \\
                            & SGP & 3.539$\pm$0.030 & 0.603$\pm$0.059 \\
                            & \hevics & 3.774$\pm$0.002 & 0.184$\pm$0.004 \\
\multirow{3}{*}{160\micron} & NGP & 3.515$\pm$0.017 & 0.389$\pm$0.055 \\
                            & SGP & 3.532$\pm$0.017 & 0.380$\pm$0.058 \\
                            & \hevics & 3.714$\pm$0.001 & 0.240$\pm$0.003 \\ 
\enddata
\tablecomments{The best-fit parameters for the relationship
between pixel noise and number of scans
(Equation~\ref{equ:pacsNoiseModel}). See Section~\ref{sec:pacs} for details.}

\end{deluxetable}

\subsection{Photometry for Point and Extended Sources}
\label{sec:pacsPhot}

The PACS PSF is not
a simple Gaussian and in fast-scan parallel mode is significantly
extended in the scan direction (Section~\ref{sec:PACSdr}), which means
that it must vary within both fields, especially between points
on the maps that are composed of different numbers of individual observations.
For this reason, the technique of maximising the signal-to-noise for point sources by convolving the images
with the PSF is not as accurate as for the SPIRE images. Instead, we use aperture photometry with a small
aperture (see below). However, if the reader does prefer to convolve the map with the PSF, for example
for detecting a faint point source, we recommend the use of our Gaussian fit to the empirical PSFs, which have
a FWHM of 11.4 and 13.7 arcsec at 100 and 160\micron, respectively. Anyone carrying out a stacking analysis should be aware
that the means of the PACS maps are not zero,
and so the mean of the map should be subtracted before proceeding with the stacking\footnote{As we noted for
SPIRE, {\it Nebuliser} produces an estimate of the
sky value at each point but this value
is not generally equal to the mean at that
point.
Therefore, stacking analyses, which sum the emission from large numbers
of sources will be sensitive to any small systematic error
in the way that {\it Nebuliser} estimates the background.
Therefore, to be safe, the mean should be subtracted from an image
before carrying out a stacking analysis.}.
Errors for a stacking analysis should be obtained from a Monte Carlo analysis in which flux densities are measured
at random points in the image.

For photometry of an unresolved source, the alternative to measuring the flux density from an image that has
been convolved with the PSF is aperture photometry with an aperture not much larger than the PSF. 
\citetalias{Valiante2016} found that the signal-to-noise peaks for an aperture with a radius of $\simeq$8 arcsec
at both wavelengths. We suggest that astronomers wishing to carry out photometry of point sources
should use this aperture, although since such a
small aperture contains only a small number of pixels, they should think carefully
about pixelization effects when using this approach.
The units of the PACS maps are Jy\,pixel$^{-1}$, so aperture
photometry can be carried out by adding up the flux density values for all 
the pixels within the aperture; there
is no need to estimate a sky value because we have already subtracted any residual
background emission using {\it Nebulizer}. 
As part of the data release, we have supplied
a file listing the EEF in the two bands out to a reference radius of
1000 arcsec (see Section~\ref{sec:PACSdr}). Both the flux densities
and the flux errors (see below) should be corrected for the fraction of the PSF that is outside the aperture
using this table. 

Photometry of sources that are expected to be extended, for example nearby galaxies,
should also be carried out by aperture photometry. 
There is no need to estimate a sky value because we have already subtracted any residual
background emission using {\it Nebulizer}. 
Both the flux densities
and the flux errors (see below) should be corrected for the fraction of the PSF that is outside the aperture
using the EEF.

\begin{figure}
  \centering
  \includegraphics[trim=20mm 12mm 4mm 4mm,clip=True, width=0.49\textwidth]{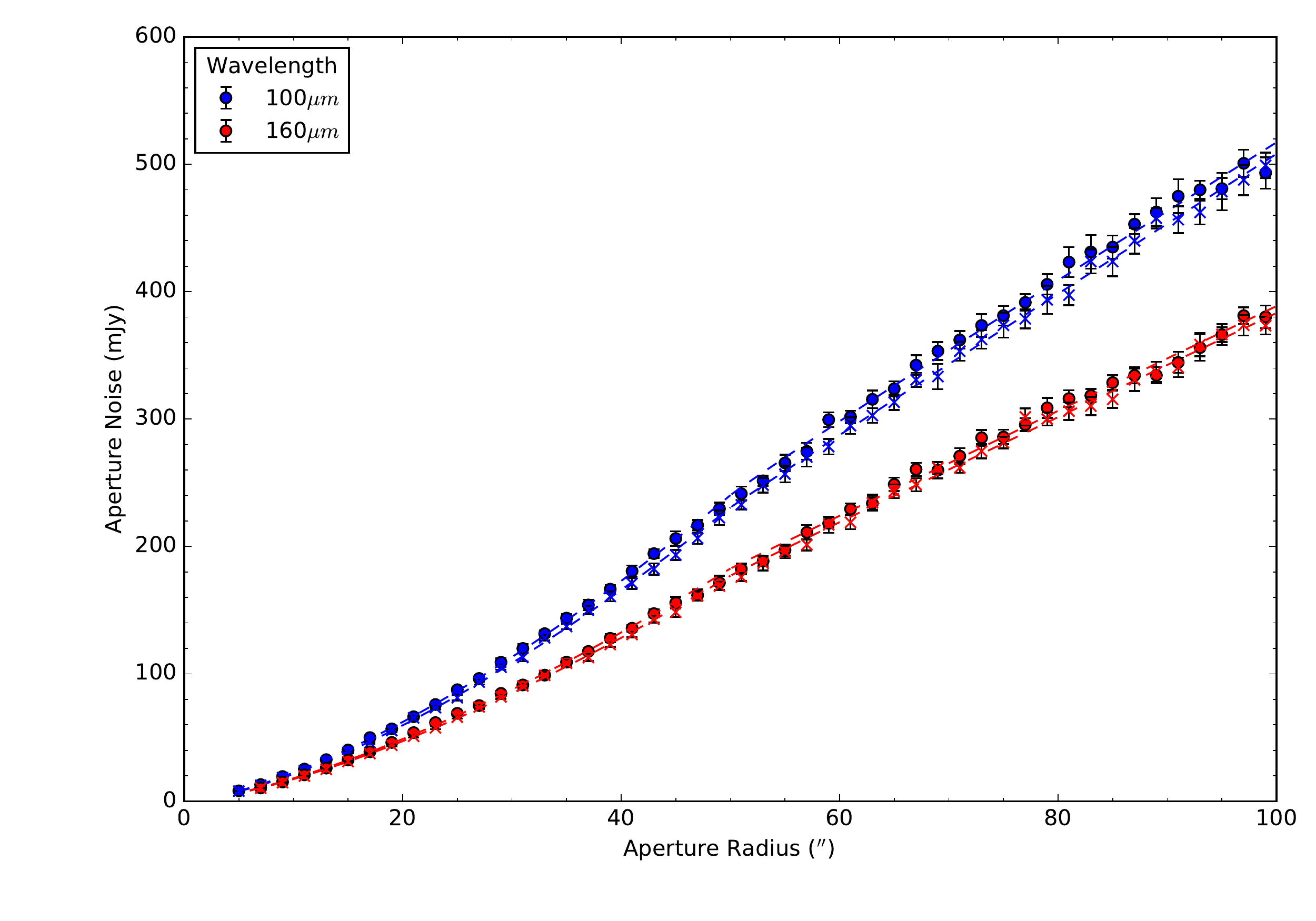}
  \figcaption{Results of the random Monte Carlo simulation for the two PACS bands, where we place 
           apertures with radius varying from approximately the
           beam size up to 100\arcsec, with 3000 apertures used at each radii. The apertures are only placed
           on regions with $\rm N_{scan} = 2$. Results from the NGP and SGP are shown by the circular and cross points, 
           respectively. The best fit models described in Section~\ref{sec:pacsPhot} are shown by the dashed lines.}
  \label{fig:PACSmonte}
\end{figure}

We have carried out a similar Monte Carlo simulation to estimate
the errors 
in the flux densities measured in aperture photometry to the one
we carried out for SPIRE
(Section~\ref{sec:spirePhot}).
For each aperture radius, we placed 3000 apertures at random positions on
the part of each image with $\rm N_{scan}=2$. We used radii ranging from approximately
the size of the beam up to 100 arcsec. The results are shown in Figure~\ref{fig:PACSmonte}.
As for SPIRE, we fit a power-law relationship (Equation~\ref{equ:apModels}) to the
results of
the Monte-Carlo simulation. The values of the best-fit parameters in this
relationship are given
in 
Table~\ref{tab:SPIREaperture}.
As for SPIRE, we found that at large radii the flux error is approximately proportional
to the aperture radius, which is the relationship expected for instrumental noise
that is not correlated between pixels.
At smaller radii, as for SPIRE, we found the flux error increases more
rapidly with radius.
We are not sure of the explanation but possibilities include cirrus emission that has
not been removed because of the large filtering scale used in {\it Nebuliser}
and
residual 1/$f$ noise not removed by the map-maker.
We have not produced maps of the PACS instrumental
noise for the data release. Instead, we have produced maps showing
the number of individual datasets ($\rm N_{scan}$) contributing to each pixel.
This map and the following equation can then be used to
obtain an
estimate of the flux density
error for any object and any aperture size:
\begin{equation}
 \scriptsize
 \sigma_{\rm ap} ({\rm mJy}) = \begin{cases}
    \sqrt{\left(A r^\alpha\right)^2 - \frac{\sigma_{\rm inst}^2 \pi}{P^2} \left(\frac{1}{2}-\frac{1}{\rm N_{scan}}\right)r^2} \\
    \hfil \hfil \hfil \text{if $r \leq 50^{\prime\prime}$}\\
    \sqrt{\left(B \left(r-50\right) ^\beta + A 50^{\alpha}\right)^2- \frac{\sigma_{\rm inst}^2 \pi}{P^2} \left( \frac{1}{2}-\frac{1}{\rm N_{scan}}\right)r^2}\\
    \hfil \hfil \hfil \text{for $r > 50^{\prime\prime}$}
 \end{cases}
 \label{equ:pacsApModels-Ncorrect}
\end{equation}
where $A$ and $B$ are best fit parameters from Table~\ref{tab:SPIREaperture}, $\sigma_{\rm inst}$ 
is the noise in mJy given in Table~\ref{tab:pacsPixNoise} and $P$ is the pixel size of the maps
in arcseconds (3\arcsec\ at 100\micron\ and 4\arcsec\ at 160\micron). As with SPIRE, if the user wishes to use
bespoke elliptical apertures a reasonable estimate can be obtained by using the estimate 
of the flux uncertainty for a circular aperture with the same area,
or alternatively, they could run their own Monte Carlo simulation on the released maps.

On top of the flux density uncertainty given by our power-law model, there is
also a fundamental calibration error. As for SPIRE, the dominant uncertainty is due to the models of
the calibration objects, in the case of PACS stars and asteroids,
which is estimated to be 5\% (PACS Calibration 
page\footnote{\url{http://herschel.esac.esa.int/twiki/bin/view/Public/PacsCalibrationWeb}}). The reproducibility
of calibration sources is measured to be $\sim$2\% \citep{Balog2014} and so, as 
in \citetalias{Valiante2016}, we add the uncertainties and
thus make the conservative assumption that the calibration uncertainty is 7\%. 
As with SPIRE, all our measurements of flux density are based on the assumption that flux density is proportional
to frequency$^{-1}$, which introduces an error if the source does not have this SED because of the width of 
the PACS bandpass filters. We refer anyone wishing to make a correction for this effect to the
PACS Color-Correction
document\footnote{\url{http://herschel.esac.esa.int/twiki/pub/Public/PacsCalibrationWeb/cc_report_v1.pdf}}.

\section{Summary}
\label{sec:conclusions}

We have presented the largest submillimeter images that have been made of the
extragalactic sky. The {\it Herschel} Astrophysical Terahertz Large Area Survey
(\hatlas) is a survey of 660 deg$^2$ in five photometric bands: 100, 160, 250, 350,
and 500\micron\ - with the PACS and SPIRE cameras. We have
described the images
of a field 180.1 deg$^2$ in size centered on the north Galactic Pole (NGP)
and of a field 317.6  deg$^2$ in size centred on the south Galactic pole. The NGP field
contains the Coma cluster. Over most of the images,
the pixel noise, including both instrumental
noise and confusion noise, is approximately 3.6, 3.5\,mJy at 100, 160\micron, and 
11.0, 11.1, and 12.3\,mJy\,beam$^{-1}$ at 250, 350, and 500\micron, 
but reaches lower values in some parts of the images.
We have described the results of an investigation of the noise properties of the images. 
We make the most precise estimate of confusion in SPIRE maps to date, finding a value 
of $3.12 \pm 0.07$, $4.13 \pm 0.02$, and $4.45 \pm 0.04$\,mJy\,beam$^{-1}$ at 250, 350, and 500\micron\ in our 
un-convolved maps. For PACS we find an estimate of confusion in our fast-parallel observations of 0.18 and 0.24\,mJy 
at 100 and 160\micron.  
The values of the confusion noise that we have measured are similar but 
not identical to the values from other {\it Herschel} surveys.
Finally, we have given recipes for using these images to carry out
photometry of objects, both objects expected to be unresolved
and those expected to be extended.

\acknowledgements

M.W.L.S and S.E. acknowledge support from the European Research Council (ERC) Forward Progress 7 (FP7)
project HELP. L.D., S.J.M. and H.G. acknowledge support from the ERC Consolidated grant cosmic dust.
L.D., S.J.M. and R.J.I. acknowledge support from the ERC Advanced Investigator grant COSMICISM.
E.V. and S.A.E. acknowledge funding  from  the  UK  Science  and  Technology  Facilities  Council 
consolidated  grant  ST/K000926/1.

We thank everyone involved with the {\it Herschel Space Observatory}.

\Hersc\ is an ESA space observatory with science instruments provided by European-led 
Principal Investigator consortia and with important participation from NASA. The
Herschel spacecraft was designed, built, tested, and launched
under a contract to ESA managed by the \textit{Herschel/Planck}
Project team by an industrial consortium under the overall
responsibility of the prime contractor Thales Alenia Space
(Cannes), and including Astrium (Friedrichshafen) responsible 
for the payload module and for system testing at spacecraft level, 
Thales Alenia Space (Turin) responsible for the
service module, and Astrium (Toulouse) responsible for the
telescope, with in excess of a hundred subcontractors.

HIPE is a joint development by the \Hersc\ Science Ground Segment Consortium,
consisting of ESA, the NASA \Hersc\ Science Center and the HIFI, PACS,
and SPIRE consortia. 

\vspace{5mm}
\facilities{\textit{Herschel}}


\software{numpy \citep{numpy}, scipy \citep{scipy}, Astropy \citep{astropy}, APLpy \citep{aplpy}, LMFIT, agpy.}




\bibliographystyle{aasjournal}
\bibliography{imagePaper} 



\appendix

\section{Data Release 2 Product Information}
\label{app:data}

In this paper we described the \Hersc\ images that form the second data release of \hatlas. All the data are available
at \url{www.h-atlas.org}. A short description of each image
released and the uses that they are optimized for are given in Table~\ref{tab:products}. 
In addition to the images listed in the table, we give the EEFs for each band, 
and a Multi-Order Coverage file \citep[MOC,][]{Fernique2014} which can be used to easily select the \hatlas\ region in other
catalogs or maps. For SPIRE we also provide the PSFs and matched-filters used.
The data release page also provides the \hatlas\ catalogs described in \citet{Maddox2017}
and \citet{Furlanetto2017}.

\movetabledown=0.25in

\begin{rotatetable}
\begin{deluxetable*}{cccp{6cm}p{6cm}}
\tablecaption{Description and Uses of the Images Released~\label{tab:products}}
\tablecolumns{5}
\tablewidth{0pt}
\tabletypesize{\scriptsize}

\tablehead{\colhead{Instrument} & \colhead{Product Identifier} & \colhead{Map Unit} & \colhead{Processing Details} & \colhead{Optimised Use Cases}}
\startdata
\multirow{7}{*}{PACS} & \multirow{3}{*}{\_BACKSUB} & \multirow{3}{*}{Jy\,pix$^{-1}$} & 
     Our released PACS images, created using the description in Section~\ref{sec:PACSdr} and shown in Figure~\ref{fig:pacsNGP} \&
     \ref{fig:pacsSGP}. These maps have been background subtracted with {\sc Nebuliser}, which removes emission on scales 
     above 300\arcsec. &
     Any flux density measurements with the PACS data should be made with these maps.\\
\cmidrule(l{10mm}r{10mm}){2-5}
   & \multirow{4}{*}{\_NSCAN}   & \multirow{4}{*}{...}      & 
     Image showing the number of individual observations that have contributed to the flux density in each pixel, as
     shown in Figure~\ref{fig:pacsNGP} \& \ref{fig:pacsSGP}. 
   & This image can be used to only select regions with a specific number of observations, and can be used together with 
     with Table~\ref{tab:pacsPixNoise} to calculate sensitivity, or with Equation~\ref{equ:pacsApModels-Ncorrect} to find
     the uncertainty on an aperture.\\
\cmidrule(l{5mm}r{5mm}){1-5}
\multirow{22}{*}{SPIRE} & \multirow{4}{*}{\_RAW}   & \multirow{4}{*}{Jy\,beam$^{-1}$} & 
     The raw SPIRE mosaic images, created using the description in Section~\ref{sec:SPIREmaps} and shown in 
     Figures~\ref{fig:spireNGP} and \ref{fig:spireSGP}. 
   & Should be used in studies requiring large-scale structure to be preserved (i.e., cirrus emission, or large-angular size galaxies). This 
     map can be used if the users wish to apply their own filtering methods.\\
\cmidrule(l{10mm}r{10mm}){2-5}
   & \multirow{3}{*}{\_BACKSUB} & \multirow{3}{*}{Jy\,beam$^{-1}$} & 
     The SPIRE maps that have been background subtracted using {\sc Nebuliser}, 
     which was set to remove emission scales greater
     than 30 pixels (equating to 3, 4 and 6\arcmin) at 250, 350, and 500\micron, respectively (see Section~\ref{sec:spirePhot}).
   & This map is recommended for performing any aperture photometry or studies of extended sources (with the exception of 
     very extended sources), and to perform a statistical stacking analysis \citep{Viero2013}.\\
\cmidrule(l{10mm}r{10mm}){2-5}
   & \multirow{3}{*}{\_FBACKSUB} & \multirow{3}{*}{Jy\,beam$^{-1}$} & 
     The point source optimized map, created by applying the matched-filter (see Section~\ref{sec:SPIREmaps}), to the background-subtracted map.
   & This map has been optimized for the detection and measurement of point sources. Any science goals investigating individual point sources should use this map.\\
\cmidrule(l{10mm}r{10mm}){2-5}
   & \multirow{4}{*}{\_NSCAN}   & \multirow{4}{*}{...}      & 
     Image showing the number of individual observations that have contributed to the flux density in each pixel, as
     shown in Figure~\ref{fig:spireNGP} and \ref{fig:spireSGP}. 
   & This image can be used to only select regions with a specific number of observations, and can be used together  
     with Table~\ref{tab:SPIREinstrumental} to calculate sensitivity, or with Equation~\ref{equ:apModels-Ncorrect} to find
     the uncertainty on a flux measurement in an aperture.\\
\cmidrule(l{10mm}r{10mm}){2-5}
   & \multirow{2}{*}{\_MASK}  & \multirow{2}{*}{...} & 
     A map showing the regions where the \hatlas\ source detection has been applied. 
   & This map can be used to see whether a particular coordinate falls within the region covered by the \hatlas\ catalogue.\\
\cmidrule(l{10mm}r{10mm}){2-5}
   & \multirow{3}{*}{\_SIGMA} & \multirow{3}{*}{Jy\,beam$^{-1}$} & The uncertainty map for our un-filtered SPIRE maps. 
     The differences between this map and the 
     default uncertainty maps produced by HIPE are described in Section~\ref{sec:SPIREmaps}. 
   & This map can be used to find the instrumental noise for any pixel on the image. This is useful as the sensitivity 
     can vary, even in regions with the same number of observations.\\
\cmidrule(l{10mm}r{10mm}){2-5}
   & \multirow{3}{*}{\_FSIGMA} & \multirow{3}{*}{Jy\,beam$^{-1}$} & 
     The uncertainty map for our matched-filtered maps (see Section~\ref{sec:SPIREmaps}). & 
     This map can be used to find the instrumental noise for any pixel on the matched-filtered map. This is useful as the sensitivity 
     can vary, even in regions with the same number of observations.\\
\enddata
\tablecomments{The images released as part of \hatlas\ Data Release 2. The file names for the products in the table all include the
field (NGP or SGP), the product identifier (column 2) and the wavelength in microns (100, 160, 250, 350, or 500).}
\end{deluxetable*}
\end{rotatetable}
\pagebreak
\section{Confusion Information}
\label{app:confusion}

In Section~\ref{sec:spirePhot} we recommended that for an individual source the confusion noise that is most appropriate to use
is from our second definition in Section~\ref{sec:SPIREconf}. In this method the confusion noise depends on the 
flux density of the source, and the relationship is shown in Figure~\ref{fig:confusion2}. 
To allow users to use the most appropriate confusion value for their source, Table~\ref{tab:confFlim} provides the confusion noise
values for each flux limit that was used to plot Figure~\ref{fig:confusion2}.

\begin{deluxetable}{c|cccc|cccc|cccc}[h]
\tablecaption{Confusion Noise versus Flux Limit\label{tab:confFlim}}
\tablecolumns{13}
\tablewidth{0pt}

\tablehead{ & \multicolumn{12}{c}{Confusion Noise (mJy\,beam$^{-1}$)} \\
           Flux & \multicolumn{4}{c}{250\micron} & \multicolumn{4}{c}{350\micron} & \multicolumn{4}{c}{500\micron}\\
           Limit & \multicolumn{2}{c}{Raw} & \multicolumn{2}{c}{Nebulised} & \multicolumn{2}{c}{Raw} & \multicolumn{2}{c}{Nebulised} & \multicolumn{2}{c}{Raw} & \multicolumn{2}{c}{Nebulised}\\
           (Jy\,beam$^{-1}$) & NGP & SGP & NGP & SGP & NGP & SGP & NGP & SGP & NGP & SGP & NGP & SGP}

\startdata
0.0100 & 1.959 & 1.676 & 1.287 & 1.575 & 2.257 & 2.030 & 1.923 & 1.708 & 2.287 & 1.795 & 1.980 & 1.964\\
0.0126 & 2.063 & 2.051 & 1.967 & 1.700 & 2.727 & 2.598 & 2.362 & 2.272 & 2.708 & 2.455 & 2.352 & 2.130\\
0.0159 & 2.722 & 2.697 & 2.281 & 2.295 & 3.265 & 3.322 & 2.995 & 2.864 & 3.153 & 3.181 & 2.927 & 2.736\\
0.0200 & 3.282 & 3.273 & 2.954 & 2.941 & 4.120 & 4.011 & 3.573 & 3.582 & 3.965 & 3.919 & 3.470 & 3.568\\
0.0252 & 4.173 & 4.160 & 3.736 & 3.752 & 4.975 & 4.989 & 4.515 & 4.506 & 4.811 & 4.765 & 4.526 & 4.296\\
0.0317 & 4.973 & 4.998 & 4.621 & 4.519 & 5.856 & 5.845 & 5.475 & 5.385 & 5.690 & 5.639 & 5.405 & 5.158\\
0.0399 & 5.521 & 5.563 & 5.206 & 5.167 & 6.408 & 6.428 & 6.050 & 5.992 & 6.212 & 6.226 & 5.894 & 5.785\\
0.0502 & 5.870 & 5.925 & 5.580 & 5.551 & 6.720 & 6.755 & 6.375 & 6.338 & 6.487 & 6.492 & 6.115 & 6.064\\
0.0632 & 6.078 & 6.146 & 5.802 & 5.781 & 6.880 & 6.912 & 6.536 & 6.505 & 6.575 & 6.604 & 6.202 & 6.180\\
0.0796 & 6.221 & 6.293 & 5.962 & 5.937 & 6.958 & 6.990 & 6.620 & 6.593 & 6.616 & 6.638 & 6.234 & 6.208\\
0.1002 & 6.334 & 6.403 & 6.071 & 6.060 & 7.015 & 7.050 & 6.670 & 6.643 & 6.637 & 6.666 & 6.259 & 6.230\\
0.1262 & 6.427 & 6.495 & 6.168 & 6.151 & 7.064 & 7.093 & 6.727 & 6.684 & 6.661 & 6.686 & 6.276 & 6.261\\
0.1589 & 6.524 & 6.575 & 6.271 & 6.239 & 7.115 & 7.120 & 6.772 & 6.719 & 6.672 & 6.709 & 6.298 & 6.275\\
0.2000 & 6.624 & 6.656 & 6.366 & 6.326 & 7.164 & 7.151 & 6.816 & 6.756 & 6.690 & 6.735 & 6.315 & 6.302\\
0.2518 & 6.726 & 6.743 & 6.476 & 6.397 & 7.209 & 7.200 & 6.863 & 6.798 & 6.709 & 6.756 & 6.330 & 6.333\\
0.3170 & 6.844 & 6.813 & 6.582 & 6.482 & 7.250 & 7.244 & 6.903 & 6.840 & 6.728 & 6.787 & 6.344 & 6.339\\
0.3991 & 6.941 & 6.912 & 6.677 & 6.597 & 7.281 & 7.292 & 6.931 & 6.883 & 6.746 & 6.801 & 6.353 & 6.380\\
0.5024 & 7.024 & 7.038 & 6.770 & 6.710 & 7.308 & 7.333 & 6.955 & 6.948 & 6.747 & 6.833 & 6.363 & 6.403\\
0.6325 & 7.106 & 7.182 & 6.819 & 6.810 & 7.347 & 7.405 & 6.983 & 6.999 & 6.753 & 6.853 & 6.372 & 6.410\\
0.7962 & 7.162 & 7.290 & 6.892 & 6.900 & 7.356 & 7.440 & 7.004 & 7.049 & 6.757 & 6.864 & 6.378 & 6.438\\
\enddata
\tablecomments{Estimates of the confusion noise in the three SPIRE bands using our second definition of confusion.
               Measurements are given for each field, and for both the raw and nebulised SPIRE maps.}
\end{deluxetable}

\end{document}